\documentclass[manuscript,screen]{acmart}

\AtBeginDocument{%
  \providecommand\BibTeX{{%
    \normalfont B\kern-0.5em{\scshape i\kern-0.25em b}\kern-0.8em\TeX}}}

\usepackage{array}
\newcolumntype{H}{>{\setbox0=\hbox\bgroup}c<{\egroup}@{}}
\newcolumntype{Z}{>{\setbox0=\hbox\bgroup}c<{\egroup}@{\hspace*{-\tabcolsep}}}

\begin{document}


\title{A Survey of Blockchain-Based Privacy Applications: An Analysis of Consent Management and Self-Sovereign Identity Approaches}

\author{Rodrigo Dutra Garcia}
\email{rgarcia@usp.br}
\orcid{0000-0003-3607-9315}
\authornotemark[1]
\affiliation{%
  \institution{Institute of Mathematics and Computer Science, University of Sao Paulo}
  \streetaddress{Avenida Trabalhador São-Carlense 400}
  \city{São Carlos}
  \state{São Paulo}
  \country{Brazil}
  \postcode{43017-6221}
}

\author{Gowri Ramachandran}
\orcid{0000-0003-3607-9315}
\affiliation{%
  \institution{Faculty of Science, Queensland University of Technology}
  \streetaddress{2 George Street, Brisbane, Queensland, 4000}
  \city{Brisbane}
    \state{Queensland}
  \country{Australia}}
\email{g.ramachandran@qut.edu.au}

\author{Kealan Dunnett}
\orcid{0000-0002-0010-1499}
\affiliation{%
  \institution{School of Computer Science, Queensland University of Technology}
  \streetaddress{2 George Street, Brisbane, Queensland, 4000}
  \city{Brisbane}
    \state{Queensland}
  \country{Australia}}
\email{kealan.dunnett@connect.qut.edu.au}

\author{Raja Jurdak}
\orcid{0000-0001-7517-0782}
\affiliation{%
  \institution{School of Computer Science, Queensland University of Technology}
  \streetaddress{2 George Street, Brisbane, Queensland, 4000}
  \city{Brisbane}
    \state{Queensland}
  \country{Australia}}
\email{r.jurdak@qut.edu.au}

\author{Caetano Ranieri}
\orcid{0000-0001-5680-9085}
\affiliation{%
  \institution{Institute of Geosciences and Exact Sciences, Sao Paulo State University}
  \streetaddress{Avenida 24 A,1515}
  \city{Rio Claro}
  \state{São Paulo}
  \country{Brazil}
  \postcode{13506-900}}
  \email{cm.ranieri@unesp.br}

\author{Bhaskar Krishnamachari}
\orcid{0000-0002-9994-9931}
\affiliation{%
  \institution{USC Viterbi School of Engineering, University of Southern California}
  \streetaddress{3650 Mcclintock Ave, Los Angeles, CA 90089}
  \city{Los Angeles}
  \state{California}
  \country{USA}
  \postcode{90089}}
  \email{bkrishna@usc.edu}

\author{Jó Ueyama}
\orcid{0000-0002-5591-3750}
\affiliation{%
  \institution{Institute of Mathematics and Computer Science, University of Sao Paulo}
  \streetaddress{Avenida Trabalhador São-Carlense 400}
  \city{São Carlos}
  \state{São Paulo}
  \country{Brazil}
  \postcode{43017-6221}}
  \email{joueyama@icmc.usp.br}

\renewcommand{\shortauthors}{Garcia R.D, et al.}

\begin{abstract}
Modern distributed applications in healthcare, supply chain, and the Internet of Things handle a large amount of data in a diverse application setting with multiple stakeholders. Such applications leverage advanced artificial intelligence (AI) and machine learning algorithms to automate business processes. The proliferation of modern AI technologies increases the data demand. However, real-world networks often include private and sensitive information of businesses, users, and other organizations. Emerging data-protection regulations such as the General Data Protection Regulation (GDPR) and the California Consumer Privacy Act (CCPA) introduce policies around collecting, storing, and managing digital data. While Blockchain technology offers transparency, auditability, and immutability for multi-stakeholder applications, it lacks inherent support for privacy. Typically, privacy support is added to a blockchain-based application by incorporating cryptographic schemes, consent mechanisms, and self-sovereign identity. This article surveys the literature on blockchain-based privacy-preserving systems and identifies the tools for protecting privacy. Besides, consent mechanisms and identity management in the context of blockchain-based systems are also analyzed. The article concludes by highlighting the list of open challenges and further research opportunities.  
\end{abstract}





\maketitle

\section{Introduction}
The rise of Internet of Things (IoT) technology in recent years has seen its adoption in a diverse range of contexts \cite{INDRAKUMARI2020163, 10.1145/3555308}. From industrial sectors to online services to healthcare, the application of this technology has enabled the collection of previously inaccessible data sources. For example, within the context of healthcare, the development of a range of smart devices that automatically monitor a user’s health (e.g., Smart Watches) in some cases facilitates the real-time collection of a range of health-related data \cite{enshaeifar2020digital}. While the collection of these data sources has several benefits, its collection also creates a number of privacy questions.

More recently, these privacy issues have been highlighted by a number of large-scale incidents that have had significant privacy implications. For example, in December of 2022, the prominent Australian Telecommunications provider Optus had a large-scale data breach that resulted in over 10 million current and past customer records being leaked \cite{oxford2022recovery}. Of particular concern was the identity-based information associated with this breach, as in some cases, customers' passports and or driver's license numbers were obtained. Subsequently, many privacy questions associated with the management of sensitive data have been raised.

Towards these privacy issues, a number of jurisdictions have introduced regulations, such as the General Data Protection Regulation (GDPR) in Europe, as a mechanism to ensure that the collection of user data is properly managed by organizations collecting it \cite{10015729}. As part of these regulations, consent management plays a significant role. More specifically, consent management describes the process of a data owner (e.g., patient or user) giving consent to how their data is used and for what purpose \cite{Kakarlapudi2021}. 

As well as user-based consent management, a number of identity management approaches have been developed in recent years to preserve the privacy of user identities \cite{9858139}. In particular, the concept of Self-Sovereign Identities (SSI) has been applied within a number of contexts to provide users with greater governance over their digital identity \cite{Feulner2022}. Similar to consent management, SSI provides users with sole ownership of their identity and, therefore, management of the personal data associated with it.


Transparency and traceability are fundamental to consent management, SSI, and machine learning-based approaches that focus on facilitating privacy. For example, when a consent management system managing access to potentially sensitive user data is considered, a lack of transparency and traceability results in the data owner not being able to monitor or track who accesses their data. As a result, the centralization of processes that facilitate these approaches, such as using a central entity to manage user identities, results in a lack of transparency and traceability. Towards this challenge, blockchain-based approaches offer a promising opportunity. In particular, blockchain's decentralized, immutable, and auditable properties can be leveraged to facilitate privacy-preserving consent management and SSI such that transparency and traceability are provided \cite{bashir2017mastering}. 

However, due to the inherent transparency associated with current blockchain technologies, the privacy of sensitive data is not inherently protected. As a result of this, several works propose privacy-enhancing mechanisms that utilize additional protocols, such as encryption \cite{our-previous-journal-work} and off-chain storage \cite{9964468}, in combination with blockchain to facilitate privacy in a variety of contexts.

Blockchain technology enables decentralization, transparency, and auditability through a distributed consensus algorithm and an immutable ledger. Applications that use this technology must reveal transaction data to consensus nodes and smart contracts to get the desired transparency and correctness benefits. Conversely, privacy focuses on hiding sensitive and personal data from other participants in the network. Let us consider a multi-stakeholder supply chain application requiring transparency and auditability. Smart contracts and consensus nodes must verify the correctness of the data before storing them on the ledger. A private transaction may reach the consensus nodes and smart contracts in encryption form, demanding decryption within the blockchain and consensus nodes. While the encrypted data could be processed (without applying decryption) as part of the blockchain consensus, it cannot guarantee correctness as it could be garbage. When blockchain technology is used for a privacy-sensitive application, it is crucial to analyze the privacy protections, consensus correctness, and auditability. Besides, the applications must integrate privacy protections within and outside the blockchain. Blockchain technology cannot provide strong privacy guarantees while offering transparency, auditability, and decentralization in a distributed and multi-stakeholder application.



Multiple studies, including \cite{rel_survey_2} and \cite{rel_survey_7}, have investigated user privacy in blockchain-based solutions, concentrating on cryptography and privacy-preserving techniques. Research by \cite{rel_survey_4} and \cite{rel_survey_5} examined application areas, including healthcare, IoT, and finance, comparing privacy protection countermeasures and smart contracts privacy platforms. \cite{rel_survey_3} and \cite{rel_survey_6} have studied privacy and security aspects in healthcare systems, discussing blockchain features to tackle security and privacy challenges. \cite{rel_survey_did_ssi_8}, \cite{rel_survey_did_ssi_9}, and \cite{10.1109/ACCESS.2022.3216643} have explored identity management concepts in self-sovereign identities. 
Despite considering privacy and self-sovereign identity, no existing study connects these aspects to examine features. To provide research opportunities in the industry, it is important to investigate solutions addressing data protection, secure collaborative learning, and self-sovereign identity management with available implementations.

This survey aims to examine the privacy-preservation mechanisms found in the literature and how such mechanisms are integrated with the blockchain by answering the following research questions:

\begin{itemize}
     \item \textbf{RQ1}: What mechanisms or strategies are used in blockchain applications to ensure privacy, consent management, and selective sharing among multiple stakeholders?
    
     \item \textbf{RQ2}: How can individuals control their identity in a decentralized infrastructure?


     \item \textbf{RQ3}: What are the existing blockchain platforms and protocols enabling user privacy and identity management?
\end{itemize}



In this survey, we provide a study on data-sharing strategies for privacy protection, consent management, and identity management in multi-stakeholder settings. We examined key contributions, use cases, implementation availability, and blockchain platform for each analysis. Furthermore, we examined existing blockchain platforms focusing on privacy protection and self-sovereign identity. This work: 

\begin{itemize}

    \item Provide an analysis of blockchain-based solutions to preserve data privacy in multi-stakeholder applications.

    \item Analyze the strategies adopted to manage data owners' (i.e., users') consent to share information with data consumers.
    
    \item Investigate solutions and protocols that focus on enabling user identity management. In particular, provides a study of self-sovereign identity works and use cases.

    \item Examine blockchain platforms and protocols that support user privacy and self-sovereign identity. Additionally, provides an analysis of open-source solutions.

    \item Provide future research opportunities in the field of privacy protection solutions.
    
\end{itemize}

The organization of this survey is as follows:
Section \ref{sec:related-works} reviews related studies. Section \ref{sec:background} provides a background of the concepts related to blockchain, consent, and identity management. Section \ref{sec:methodology} outlines the methodology adopted in this study. Section \ref{sec:data-sharing-in-multi-stakeholders} analyzes blockchain-based solutions for providing privacy, consent, and identity management. Sections \ref{sec:platforms-with-focus-on-privacy}, \ref{sec:identity-management-platforms}. Section \ref{sec:research-opportunities} identifies the opportunities for further research in this field, and Section \ref{sec:conclusion} concludes the survey and summarizes the key findings.

\section{Related Work}
\label{sec:related-works}

Some work analyzed different protocols to provide user privacy in blockchain-based solutions. For instance, \citeauthor{rel_survey_2} \cite{rel_survey_2} analyzed some privacy threats and protection mechanisms in blockchain-based applications. In particular, the authors introduced cryptography techniques such as mixing services, ring signatures, and non-interactive zero-knowledge proofs. Similarly, \citeauthor{rel_survey_7} \cite{rel_survey_7} reviewed existing privacy-preserving techniques for blockchain applications. In particular, the authors divided the topic of privacy into the following categories: Smart Contract/Key Management, Identity Data Anonymization, Transaction Data Anonymization, and On-chain data protection. Also, the survey presents the SSI concepts and applications for contributions up to 2019.

\citeauthor{rel_survey_4} \cite{rel_survey_4} evaluated contributions in different application areas up to 2018. The authors analyzed several blockchain use cases, such as healthcare, the Internet of Things, and financial use cases, and showed some challenges regarding security and privacy. Similarly, \citeauthor{rel_survey_5} \cite{rel_survey_5} analyzed the possible threats to privacy in blockchain applications. In particular, the survey compares different countermeasures and methods for privacy protection, such as non-interactive zero-knowledge proof and mixing. Also, the authors examined some platforms to provide privacy in smart contracts.

\citeauthor{rel_survey_3} \cite{rel_survey_3} analyzed privacy and security aspects in healthcare systems. The survey summarizes some technologies used for privacy preservation in data sharing among different sources, such as hospitals and laboratories. Moreover, the authors discussed some smart contract implementation details, such as access control and encryption mechanisms to store Electronic health records (EHR). Likewise, \citeauthor{rel_survey_6} \cite{rel_survey_6} outlined how blockchain features can improve some security and privacy challenges. In particular, the survey presents blockchain concepts such as cryptography, consensus mechanisms, and security threats. \citeauthor{rel_survey_1} \cite{rel_survey_1} investigated security and privacy in blockchain solutions divided into three categories: (1) attack and defense methods in blockchain-based mining, (2) network and smart contract attacks and (3) blockchain privacy issues such as identity privacy and transaction data. 

\citeauthor{rel_survey_did_ssi_8} \cite{rel_survey_did_ssi_8} introduced the concepts of identity management (IdM) and presented an analysis of blockchain-based solutions for self-sovereign identities. In particular, the authors explored how blockchain features can be used for identity IdM and presented some platforms for identity-based solution development. Similarly, \citeauthor{rel_survey_did_ssi_9} \cite{rel_survey_did_ssi_9} examined some requirements and functionalities for SSI technologies. Moreover, the authors studied some blockchain works focusing on identity management. \citeauthor{10.1109/ACCESS.2022.3216643} \cite{10.1109/ACCESS.2022.3216643} describes five IdM components: authentication, integrity, privacy, trust, and simplicity. Additionally, the authors provided a security analysis to identify threats that could harm blockchain-based IdM systems.


Despite the blockchain-based work's contributions, they do not provide an analysis combining privacy, consent, and identity management. In this work, we analyze privacy-protecting solutions and platforms in decentralized settings. In particular, we examined privacy protocols, consent, and identity management in multi-stakeholder systems. Table \ref{tab:related-surveys} summarizes existing surveys on privacy preservation with blockchain and their limitations.


\begin{table*}[h]
\centering
\caption{Comparison of Related Surveys with This Study}
\label{tab:related-surveys}

\resizebox{1\textwidth}{!}{

\begin{tabular}{lccccc}
\hline
\multicolumn{6}{c}{\textbf{Contributions}}                                                                                                                                                                                                                                                                                                                                                                                                                            \\ \hline
\multicolumn{1}{c}{\textbf{\begin{tabular}[c]{@{}c@{}}Related \\ Work\end{tabular}}} & \textbf{\begin{tabular}[c]{@{}c@{}}Publication\\ Year\end{tabular}} & \textbf{\begin{tabular}[c]{@{}c@{}}Privacy\\ Support\end{tabular}} & \textbf{\begin{tabular}[c]{@{}c@{}}Consent\\ Management\end{tabular}} & \textbf{\begin{tabular}[c]{@{}c@{}}Self-Sovereign\\ Identity\end{tabular}} & \textbf{\begin{tabular}[c]{@{}c@{}}Implementation\\ Availability\end{tabular}} \\ \hline
\citeauthor{rel_survey_2}\cite{rel_survey_2}                                         & 2019                                                                & \checkmark                                                         &                                                                       &                                                                            &                                                                                \\ \hline
\citeauthor{rel_survey_4}\cite{rel_survey_4}                                         & 2019                                                                & \checkmark                                                         &                                                                       &                                                                            &                                                                                \\ \hline
\citeauthor{rel_survey_7}\cite{rel_survey_7}                                         & 2019                                                                & \checkmark                                                         &                                                                       & \checkmark                                                                 &                                                                                \\ \hline
\citeauthor{rel_survey_3}\cite{rel_survey_3}                                         & 2020                                                                & \checkmark                                                         &                                                                       &                                                                            &                                                                                \\ \hline
\citeauthor{rel_survey_5}\cite{rel_survey_5}                                         & 2020                                                                & \checkmark                                                         &                                                                       &                                                                            &                                                                                \\ \hline
\citeauthor{rel_survey_6}\cite{rel_survey_6}                                         & 2021                                                                & \checkmark                                                         &                                                                       &                                                                            &                                                                                \\ \hline
\citeauthor{rel_survey_1}\cite{rel_survey_1}                                         & 2022                                                                & \checkmark                                                         &                                                                       &                                                                            &                                                                                \\ \hline
\citeauthor{rel_survey_did_ssi_8}\cite{rel_survey_did_ssi_8}                         & 2022                                                                & \checkmark                                                         &                                                                       & \checkmark                                                                 &                                                                                \\ \hline
\citeauthor{rel_survey_did_ssi_9}\cite{rel_survey_did_ssi_9}                         & 2022                                                                & \checkmark                                                         &                                                                       & \checkmark                                                                 &                                                                                \\ \hline
\citeauthor{10.1109/ACCESS.2022.3216643} \cite{10.1109/ACCESS.2022.3216643}          & 2022                                                                & \checkmark                                                         &                                                                       & \checkmark                                                                 &                                                                                \\ \hline
\textbf{This Survey}                                                                 & -                                                                   & \textbf{\checkmark}                                                & \textbf{\checkmark}                                                   & \textbf{\checkmark}                                                        & \textbf{\checkmark}                                                            \\ \hline
\end{tabular}

}

\end{table*}

\section{Background}
\label{sec:background}

\begin{figure*}[h]
\centering
    \centering
    \includegraphics[scale=0.5]{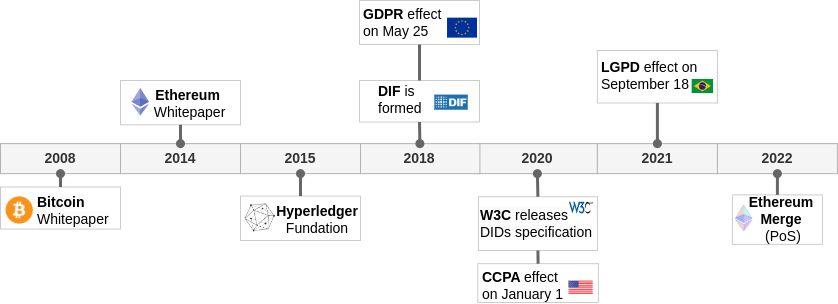}
    \caption{A timeline of some significant events in the context of blockchain, decentralized identity, and privacy regulations from 2008 until 2022}
    \label{fig:time-line-background}
\end{figure*}

This section describes blockchain technology concepts and privacy challenges. Additionally, we present the concepts of consent and identity management models. Figure \ref{fig:time-line-background} shows a timeline of significant events in the context of blockchain and privacy regulations.

\subsection{Blockchain}
\label{sec:background-blockchain}
Initially emerged with Bitcoin \cite{bitcoin-paper}, blockchain is a technology with several benefits for different applications such as industry, finance, healthcare, and the Internet of Things. In addition, blockchain enables opportunities in cybersecurity, supply chain, and machine learning research \cite{2021-archiving-cybersecurity-blockchain-survey,2022-iiot-supply-chain-survey, 2022-blockchain-federated-ml-survey}. Unlike centralized architectures, blockchain operates decentralized via a peer-to-peer (P2P) network, i.e., without a central point of failure and control. The data sent via transaction are stored in the blockchain from consensus criteria among participant nodes.

\begin{figure*}[ht!]
    \centering
    \includegraphics[scale=0.35]{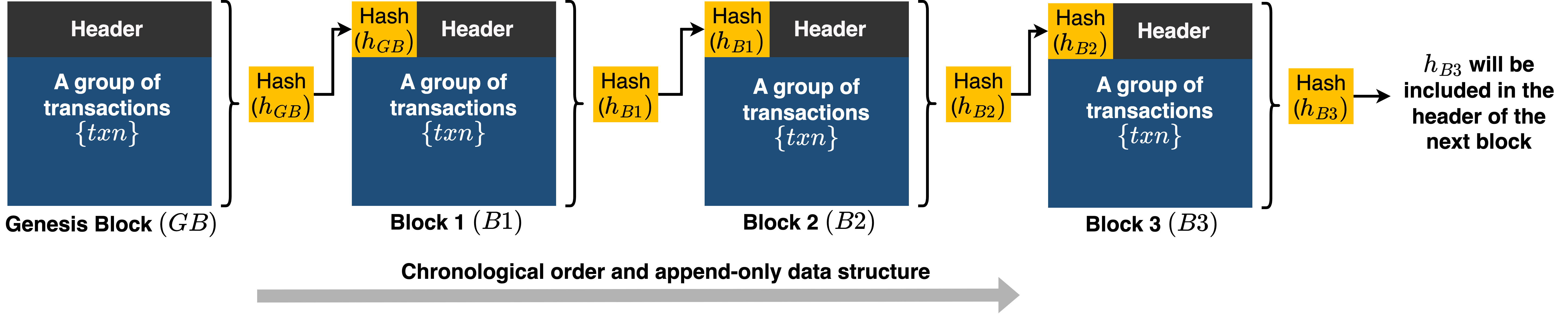}
    \caption{Blockchain data structure: blocks connected in chronological order in an append-only manner using a cryptographic hash}
    \label{fig:blockchain-structure}
\end{figure*}

Blockchain uses cryptography algorithms to secure transactions and protect the integrity of the records. In terms of data structure, the records are organized into blocks connected in a chain-like structure. In general, blocks are divided into a header containing some cryptographic fields, such as the previous block's hash, and a field containing transaction data. Figure \ref{fig:blockchain-structure} shows the basic blockchain structure. The first block created is called the genesis block, and the following transactions will be stored in an append-only manner.

In general, blockchain has the following features:

\begin{itemize}
    \item \textbf{Decentralization}: The system operations, including processing transactions, verification, and storage, follow a decentralized architecture. Centralized systems use a single physical computing node managed by a single organization for its operations without offering transparency and auditability. On the other hand, Blockchain relies on multiple physical computing nodes for its operations, using a combination of a consensus protocol, cryptographic algorithms, and an immutable storage mechanism. These physical compute nodes are often owned and managed by multiple organizations or stakeholders. 
    \item \textbf{Cryptography}: Blockchain technology uses public-key cryptography to manage transactions and blocks, which means each blockchain user creates a key pair: public and private keys. When a sender submits a transaction, they sign it using their private key. The transaction receiver can verify the sender by checking its authenticity using its public key.
    \item \textbf{Immutability}: Blockchain transactions get stored in a tamper-proof ledger, meaning once a transaction is written to a blockchain ledger, it cannot be modified. Because of this property, the blockchain ledger is called the ``write-once" ledger. Note that a network of compute nodes store and manage the blockchain ledger. Any changes to the ledger require the approval of most nodes (determined by the consensus algorithm). In a blockchain network with 1000 nodes, more than 500 nodes must approve a change to the ledger (when its consensus protocol requires 51\% approval for changes). Besides, such changes disrupt a block that holds the transactions and the other blocks in the chain because the blockchain comprises a collection of blocks. Each block is connected to its neighbouring blocks, which means any modifications to the past blocks would break the integrity of the entire blockchain. 
    \item \textbf{Traceability}: Transactions get stored in a chain of blocks in the blockchain ledger following the order in which they were executed, allowing users to trace the evolution of a given asset. For example, supply chain applications can reliably track an asset's journey by checking the transactions stored in the blockchain ledger. 
\end{itemize}

\subsubsection{The interaction process with blockchain network}

The interaction process between a user and a blockchain network starts with transaction creation. A transaction is an object created by the user application to send data to the blockchain. When a user sends the transaction object, the network verifies the transaction's validity by checking if the user has the necessary permissions and resources (such as cryptocurrency balances) to execute the transaction. At block creation time, the network will select a group of valid transactions, including those in a block, and broadcast the new block to the network. Following a consensus algorithm, the block will be securely added to the blockchain's tamper-proof ledger.

The consensus algorithm ensures that all nodes in the network agree with the block content, i.e., a group of transactions. \citeauthor{2022-survey-consensus-algorithm} \cite{2022-survey-consensus-algorithm} reviewed several consensus protocols such as Proof-of-Work (PoW), Proof-of-Stake (PoS), Delegated Proof-of-Stake (DPoS), and other mechanisms of Proof-of-X.

\subsubsection{Public and Permissioned Blockchains}
Blockchain platforms and their application deployments are broadly classified into public and private blockchains. The public blockchains allow any public member to join the network and participate in the blockchain mining, consensus, and ledger management processes. Bitcoin and Ethereum are examples of public blockchains. Users may have to pay special attention to data privacy as these blockchains make the transactions and blocks available to the public, offering maximum transparency and auditability. However, they don't have any built-in intelligence to distinguish between private and public data. Through the public keys and other identity management schemes, users may be able to remain anonymous, but the literature reports deanonymisation attacks in Bitcoin~\cite{Reid2013}.

Permissioned blockchains, on the other hand, operate in a closed and private network involving known organizations and authorized computing nodes. These types of blockchains are similar in terms of the use of consensus protocols, cryptographic techniques, and immutable ledgers, but they don't allow unauthorized nodes or organizations to join the blockchain network. Although this type of blockchain platform operates in a private and trusted environment, it still requires privacy protection as some organizations may not be willing to disclose their data to other competing organizations in the network for business and intellectual property reasons. Therefore, privacy protection is necessary for permissioned blockchains as well.

\subsection{Smart Contracts}

A contract is an agreement about a particular context among different parties. The entities typically trust a third party or mediator to encourage the agreement and resolve conflicts. In contrast, smart contracts first proposed by \citeauthor{szaboo-formalizing} \cite{szaboo-formalizing} are a piece of software running on the blockchain in a decentralized manner. In other words, smart contracts are immutable rules created via a particular programming language, enabling automation without a trusted third party \cite{2022-survey-smart-contract-formal}. 

The critical difference between a contract and a smart contract is security and immutability. Instead of a physical document, such as a paper-based contract, a smart contract is created digitally using lines of code and stored in the blockchain. Once a smart contract is deployed on the network, the business rules cannot be altered or deleted. In general, smart contracts can be used for process automation in sectors such as financial services, supply chains, the Internet of things, healthcare, and the government \cite{2022-survey-smart-contracts}.





    

\subsection{Privacy in Data Sharing}
\label{sec:privacy-data-sharing}

Privacy is the ability to choose what personal information to share with others. In recent years, the collection of personal data from a diverse range of sources facilitated through the adoption of new technologies, has become commonplace. Users or devices generate data in financial transactions, medical applications, the Internet of Things deployments, and supply chain systems. As a result, this data has significant value to organizations given its potential to be collected in large values, thus enabling its analysis to offer new products and improve user experience. However, from a user-based perspective, consideration of the security of these data assets must be preserved.

As a result of this trend toward mass data collection, protecting personal information has become a significant concern in big data. Individuals' data is often collected and stored by various third-party companies and organizations when using online services. From the Internet of Things perspective, large amounts of personal data are collected and shared \cite{2021-bg-data-privacy-iot, 2022-bg-sec-privacy-survey}. Some issues related to data privacy include:

 \begin{itemize}
     \item \textbf{Data collection and sharing:}  IoT devices collect data, including sensitive information such as location, heart rate, and other vital parameters. This data may be shared with a third party without user permission.

     \item \textbf{Lack of control and transparency:}  Users may have limited control of their data and lack transparency during data collection and usage.

     \item \textbf{Security:} Individuals may face several problems without user data security. These issues include identity theft and fraud. A malicious user can use personal information to impersonate an individual and steal their identity to commit fraud in services such as banking and healthcare. 

 \end{itemize}

Industries must implement adequate data protection and security measures to address these problems. In addition, users should know the kind of data their IoT devices collect and how it is used.

\textbf{Privacy protection in blockchain:} Maintaining the privacy of individuals or organizations involved in transactions associated with blockchain is a significant challenge. At the same time, blockchain decentralization provides transparency, i.e., all transactions are publicly visible and verifiable, as presented in Section \ref{sec:background-blockchain}. As \citeauthor{bitcoin-paper} \cite{bitcoin-paper} introduced, a transaction object has a sender address, receiver address, and data (e.g., amount of coins). To avoid linking in different transactions, \citeauthor{bitcoin-paper} explains using different key pairs for each transaction in the Privacy section of Bitcoin paper. \citeauthor{2018-anony-priv-bitcoin} \cite{2018-anony-priv-bitcoin} reviewed anonymity and privacy issues in the Bitcoin network, such as Linking Bitcoin Addresses and Mapping Bitcoin Addresses to IP Addresses. Additionally, \citeauthor{2022-sok-private-bitcoin} \cite{2022-sok-private-bitcoin} studied various privacy-enhancing techniques for the Bitcoin network, including threshold signatures, atomic swap, CoinJoin, and centralized mixing. In the Internet of Things context, \citeauthor{10.1109/JIOT.2022.3194671} examined how encrypted blockchain data can reveal people's activities. Blockchain-based solutions must provide mechanisms to protect privacy when the system handles sensitive data. In applications using personal data, the user sends a transaction containing sensitive information to the network, and without an appropriate protection scheme, the participant's nodes have access to the records, violating data privacy.



\subsection{Privacy Regulations}
Digital applications are increasingly handling sensitive data belonging to users and organizations. Noticing this trend, several countries have introduced privacy regulations, which the applications must comply with. This section presents some prominent privacy regulations to consider in developing systems, including GDPR, CCPA, HIPAA, and LGPD.

\subsubsection{General Data Protection Regulation (GDPR)}
General Data Protection Regulation (GDPR) took effect on May 25, 2018, in which organizations must adopt mechanisms for the data protection of European Union (EU) residents. Regulation impacts the management of personal data, and existing systems must adopt mechanisms for data protection. According to GDPR, personal data refers to information identifying an individual, such as name, email, telephone, and IP address \cite{GDPR}. It is one of the most advanced data protection and privacy regulations.

\subsubsection{California Consumer Privacy Act (CCPA)}

Effective January 1, 2020, CCPA is a privacy law in the U.S. state of California that gives consumers the right to know what personal information the services collected about them. Moreover, the users can request to remove personal information \cite{CCPA}. Additionally, the California Privacy Rights Act (CPRA), effective January 1, 2023, enhances CCPA by giving consumers further control over their sensitive personal data and imposing stricter obligations on businesses regarding data usage and transparency.

\subsubsection{Health Insurance Portability and Accountability Act (HIPAA)}

The federal law of 1996 created standards to protect personal health information (PHI) from being shared or disclosed without authorization \cite{HIPAA}.

\subsubsection{Brazilian General Data Protection Law (LGPD)}

LGPD is similar to GDPR in the European Union, and in terms of implementation, applies to Brazilian citizens and residents \cite{lgpd, LGPD-conference-paper}.

These regulations protect individual rights by allowing the user to control their personal information. Additionally, set the organization's responsibilities for data handling practices.

\subsection{Consent Management}

\begin{figure}[ht!]
    \centering
    \includegraphics[scale=0.5]{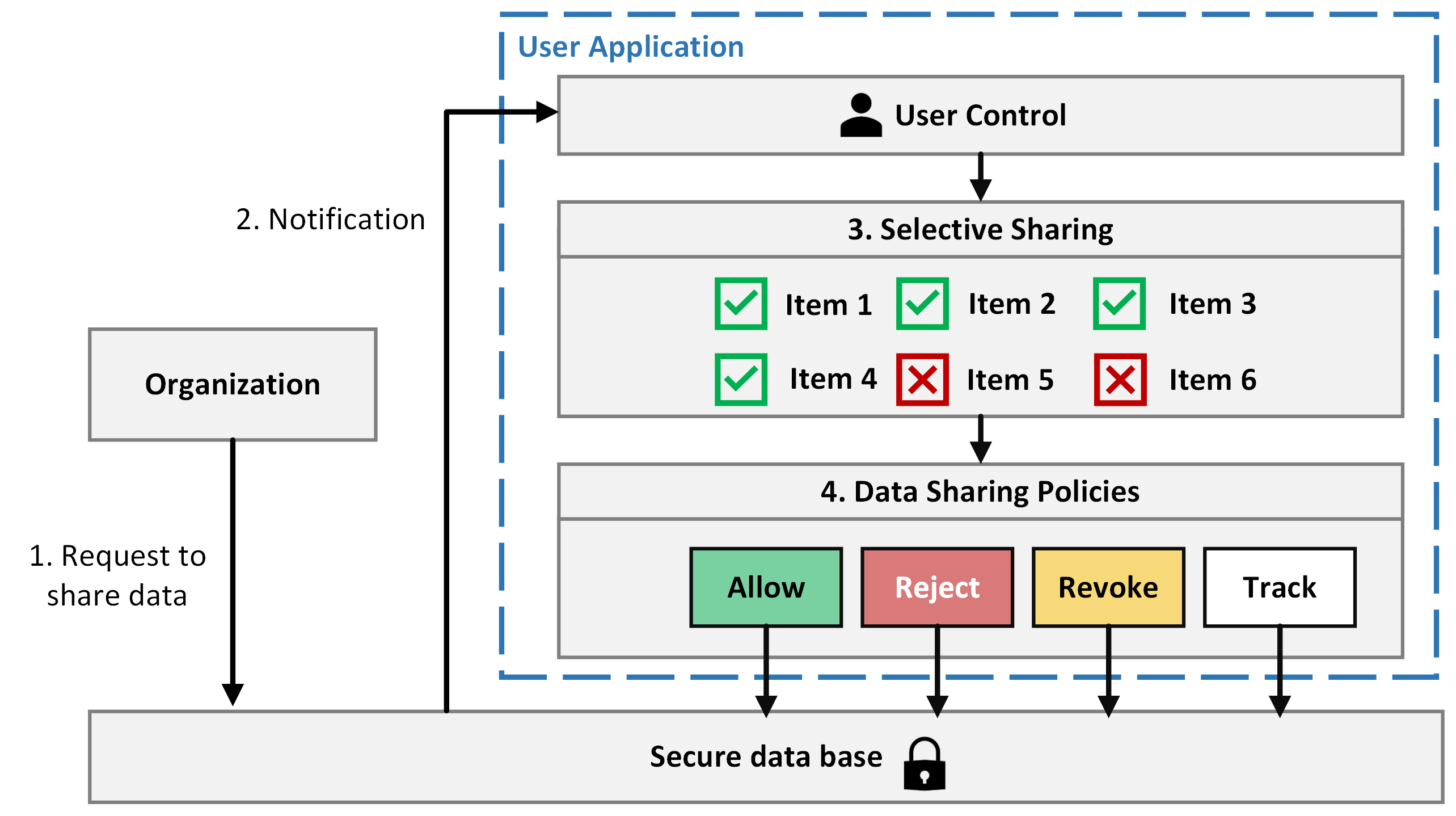}
    \caption{The flow of consent requests, user control, and selective sharing}
    \label{fig:consent-request}
\end{figure}

Consent mechanisms are policies to manage personally identifiable information (PII), enabling a user to have complete control over data sharing, including accepting, rejecting, and revoking consent to access data. Through a consent mechanism, the user will be able to track and control the use of their data by third parties. To ensure the integrity of user permissions, consent policy information must be stored in a secure location to prevent tampering and privacy violations. Consent management practice is essential to comply with data protection and privacy regulations, such as the GDPR in the European Union. Figure \ref{fig:consent-request} shows a request for data sharing that the user can allow or reject. In case of permissions are already granted, the user will be able to track the use of their data and revoke access. Some requirements in consent systems are the following:


\begin{itemize}
    \item \textbf{The integrity of access policies:} The consistency and reliability in defining and enforcing who can access data under what conditions.

    \item \textbf{User control and transparency:} Users can choose which information they share, how it is utilized, and who has permission to see it.
    
    \item \textbf{Track data usage:} The user must monitor how other parties access, process, and share data.
\end{itemize}

Consent management is generally collected using digital forms and reveals to the user why the data are necessary, following a particular organization policy \cite{Kakarlapudi2021}. When using digital services, the user is allowed to share personal information on a website. For example, patients control which data to share with other institutions in healthcare applications. In the context of federated learning, consent management plays a crucial role in protecting privacy during the learning process across multiple devices.


\subsection{Self-sovereign Identity (SSI)}
\label{sec:background-ssi}

Physical identity documents are widely adopted around the world. Some authorities are responsible for issuing physical records such as government IDs, driving licenses, and passports. There are some challenges to managing digital credentials using different technologies. It is necessary to use a standard format to verify the credential source and integrity \cite{9858139}. From a technical point of view, using standard encryption algorithms to build digital identity systems is necessary. The World Wide Web Consortium (W3C)\footnote{\url{https://www.w3.org/}}, a community focused on web standards, proposed an identity standardization to enable verifiable claims in different applications, such as healthcare and education qualifications. Regarding identity systems, there are three architectures to manage digital identities: centralized, federated, and decentralized.

\subsubsection{Centralized Model} Generally, centralized models are one of the most used architectures to manage users' credentials and personal data. The data, such as username, password, and personal information, is stored on a central database server to enable the user to access services. Therefore, the client user must allow the centralized organization to hold their information as illustrated in Figure \ref{fig:bg-centralized-id-management}. A centralized approach has drawbacks: lack of transparency, portability, and complex account management.

    


    

\subsubsection{Federated Model} 
\begin{figure}[ht!]
    \centering
    \includegraphics[scale=0.32]{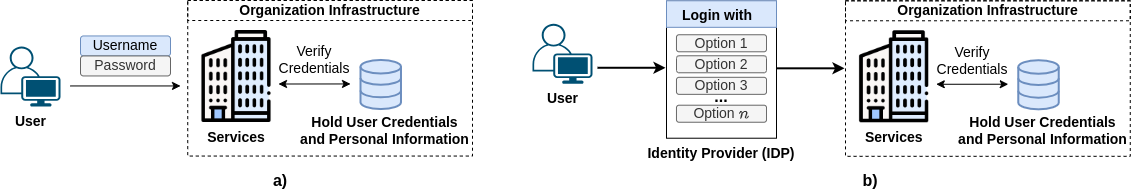}
    \caption{a) Centralized Identity Management and b) Federated Identity Management (FIM) models}
    \label{fig:bg-centralized-id-management}
\end{figure}

As illustrated in Figure \ref{fig:bg-centralized-id-management}, the federated identity management (FIM) includes an identity provider (IDP) to access some feature in a website or service. The federated identity model implements protocols such as OpenID Connect to enable users to select social media and use it as a login in the marketplace, travel agency, and other applications. There are limitations in the federated model, including privacy concerns, lack of support for some applications, and intermediation between users and services.




    


\subsubsection{Decentralized Model}

\begin{figure}[h]
    \centering
    \includegraphics[scale=0.5]{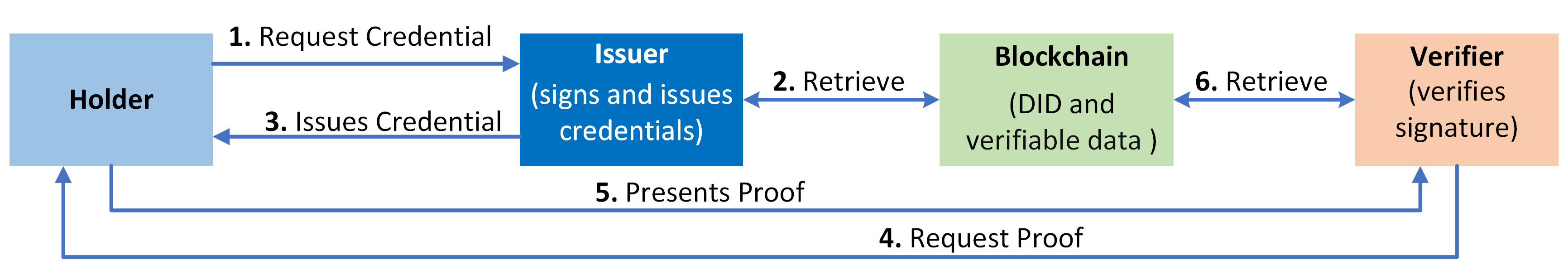}
    \caption{The actors and roles in decentralized identity management}
    \label{fig:bg-decentralized-id-management}
\end{figure}

The decentralized model doesn't rely on either centralized or federated identity systems. Instead, the user owns and controls digital identity through a digital wallet and uses the credentials to authenticate in different services \cite{9858139}. Figure \ref{fig:bg-decentralized-id-management} illustrates the decentralized model. In general, the digital identity ecosystem has three actors defined by W3C with the following attributions:

\begin{itemize}
    \item \textbf{Issuer:} create claims about the subject. For example, the government issues a digital document.
    
    \item \textbf{Holders/Owners:} hold their credentials using a digital wallet and present the proof of claims when requested.
    
    \item \textbf{Verifier:} request and verify proofs of claim from holders. In particular, the verification process uses cryptographic signatures.

\end{itemize}

In terms of identity management, the decentralized approach presents the following benefits:

\begin{itemize}
    \item Provide users with data control for different institutions such as healthcare, banking, and government services. Consequently, improve user privacy and control over their personal information.
    
    \item Enable organizations to digitally verify the user's identity without storing some sensitive data and offer better services.

\end{itemize}

However, there are some challenges to building decentralized identity systems \cite{10.1109/ACCESS.2022.3216643}, including:

\begin{itemize}
    \item Scalability to handle transactions and storage when the system has an increased number of users.
    \item The regulation of decentralized infrastructure cannot be accepted or approved by all governments, institutions, or companies, which can limit their adoption.
\end{itemize}

\subsubsection{Selective Disclosure}

Allow the entity owner to choose a piece of the data to be shared. For example, the holder has a credential issued by the government and, in a particular context, needs to reveal only the birth date instead of all credential items. The holder will select the birth date field in the application and create a verifiable derived proof for a verifier. Selective sharing is possible using digital signatures, particularly zero-knowledge proofs (ZKP). The proof will be verified using the issuer's public key on the verifier's side. Note that this scheme allows users to selectively withhold sensitive information from specific organizations. In this section, we have provided background to the main building blocks that allow privacy-preserving blockchain solutions.

\section{Survey Methodology}
\label{sec:methodology}

\begin{figure*}[ht!]
    \centering
    \includegraphics[scale=0.5]{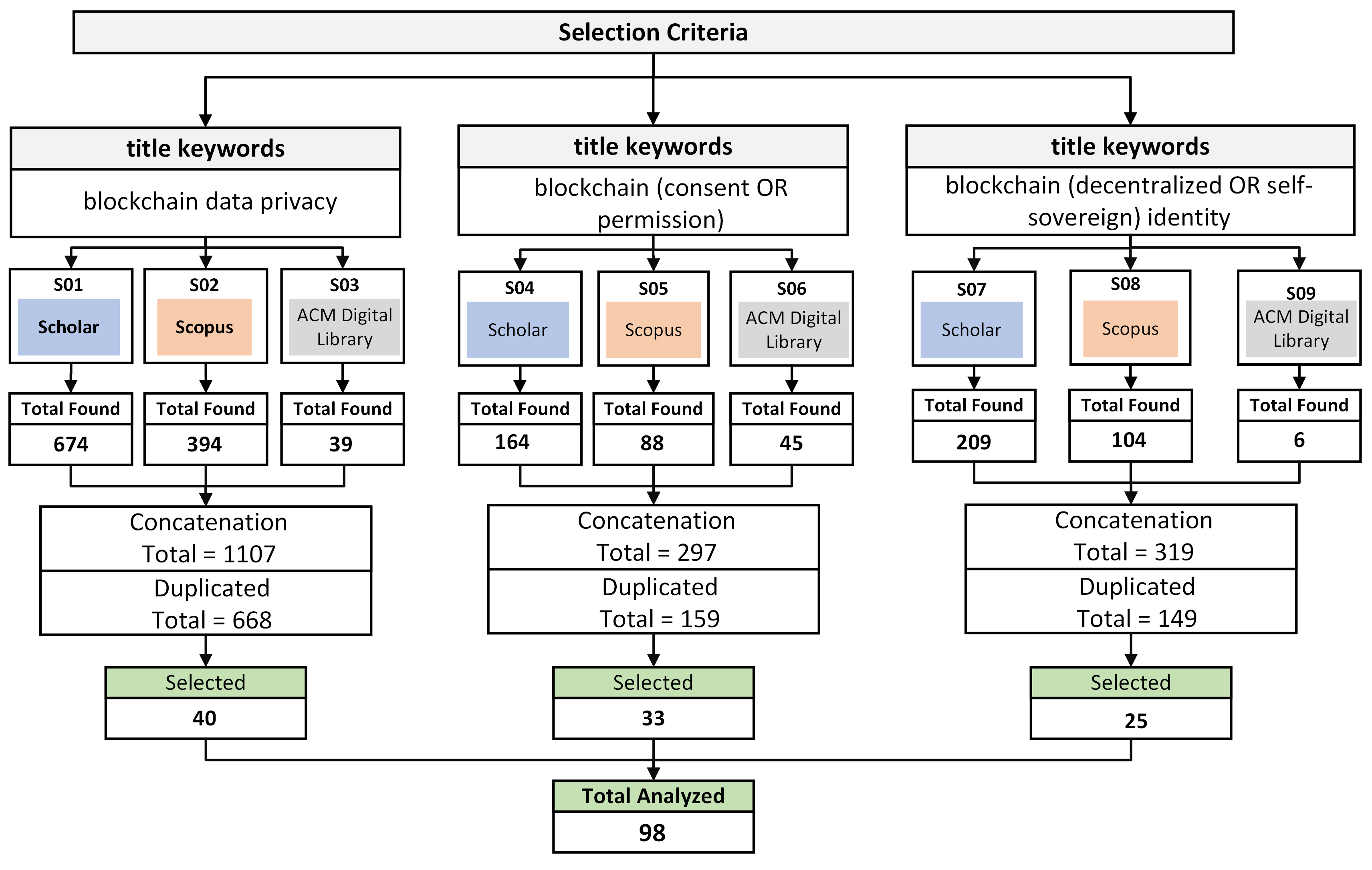}
    \caption{Relevant publications aligned with the objectives of this survey obtained from Google Scholar, Scopus, and ACM Digital Library}
    \label{fig:selection-criteria}
\end{figure*}

This survey analyzes the key contributions and recent applications in privacy, consent management, and self-sovereign identity using blockchain technology. We used the selection criteria in Figure \ref{fig:selection-criteria} to search papers in Google Scholar, Scopus, and ACM Digital Library. 

\begin{itemize}

    \item For works emphasizing privacy, we use the following search string (S):
    \begin{itemize}
        \item \textbf{Google Scholar (S01):} \textit{“allintitle: blockchain data privacy”}

        \item \textbf{Scopus (S02):} \textit{“(TITLE ( blockchain )  AND  TITLE ( data )  AND  TITLE ( privacy ) )”}

        \item \textbf{ACM Digital Library (S03):} \textit{“[Title: blockchain] AND [Title: data] AND [Title: privacy]”}

    \end{itemize}

    \item In the case of contributions in the field of consent and permission, we used the following search string:

        \begin{itemize}
        \item \textbf{Google Scholar (S03):} \textit{“allintitle: blockchain (consent OR permission)”}

        \item \textbf{Scopus (S04):} \textit{“(TITLE ( blockchain ) AND (TITLE ( consent ) OR TITLE ( permission ) ) )”}

        \item \textbf{ACM Digital Library (S05):} \textit{“[Title: blockchain] AND [[Title: consent] OR [Title: permission]]”}

    \end{itemize}




    \item For contributions in self-sovereign identity, we adopted the following search string:
    \begin{itemize}
        \item \textbf{Google Scholar (S07):} \textit{“allintitle: blockchain (decentralized OR self-sovereign) identity”}

        \item \textbf{Scopus (S08):} \textit{“( TITLE ( blockchain )  AND  TITLE ( ( decentralized  OR  self-sovereign ) )  AND  TITLE ( identity ) )”}

        \item \textbf{ACM Digital Library (S09):} \textit{“ [Title: blockchain] AND [[Title: decentralized] OR [Title: self-sovereign]] AND [Title: identity] ”}

    \end{itemize}

\end{itemize}
All keywords were applied in the title of the works, and we did not restrict publication years. In total, we selected \textbf{40} works with an emphasis on user privacy, \textbf{33} with contributions on user consent, and \textbf{25} for decentralized and self-sovereign identity. In total, \textbf{98} works were selected and will be analyzed in the next Section.

\textbf{Availability of the dataset:} all works analyzed in this survey are available on GitHub \cite{our-github-survey}.


\section{Data-sharing in multi-stakeholders applications}
\label{sec:data-sharing-in-multi-stakeholders}

This section analyzes blockchain-based contributions focusing on privacy, consent management, and self-sovereign identity approaches. Firstly, we discuss the importance of the studies of these areas jointly for industrial applications. Second, we examine the research works with their respective technological and strategic contributions.

\subsection{Why blockchain, privacy, consent, and identity management?}

Blockchain alone cannot guarantee privacy as it does not have built-in intelligence to distinguish between private and public data. Applications must, therefore, integrate privacy-protection mechanisms to protect sensitive and confidential data. We survey the blockchain-based privacy preservation approaches in Section~\ref{sec:privacy_and_blockchain} and identify the core privacy protection techniques for blockchain-based applications. 

Applications, in some cases, may include the data owner in the data-sharing pipeline to decide with whom and for what purpose the data can be shared. These types of applications have a consent mechanism through which the user would express policies for data disclosure. Section~\ref{sec:consent} surveys and analyses the consent management approaches in the context of blockchain.


Lastly, identity management platforms help applications track users and their identities in a digital application. A robust identity management framework in a blockchain-based application allows the data owners, consumers, processors, and regulators to enforce privacy policies carefully. Furthermore, a secure identity-management framework can potentially facilitate the implementation of reputational systems. Each entity in a data-driven application would receive a rating based on its past performance, providing a trust indicator for the data owner. As a result, a data owner may be willing to disclose his private data if the data consumer has a good reputation for handling or processing the user's data without compromising privacy and security. Several blockchain-based identity management systems are present in the literature. Section~\ref{sec:identity-management-platforms} reviews and synthesizes the identity management frameworks as a crucial building block in a privacy-preserving blockchain application.

\subsection{Blockchain meets Privacy}
\label{sec:privacy_and_blockchain}
This section reviews the approaches used to preserve data privacy in multi-stakeholder applications. We analyzed these approaches and grouped our results into four main use case categories based on the results: healthcare and personal data, the Internet of Things, smart grids, and alternative applications such as government systems. Tables \ref{tab:privacy-works} and \ref{tab:privacy-works-2} provide an overview of the analyzed works, including their publication year, the use case investigated, their key contributions, the blockchain platform used for prototype implementation, and whether the implemented software is available. 

As presented in Tables \ref{tab:privacy-works} and \ref{tab:privacy-works-2}, healthcare is the most common use case, followed by the Internet of Things and the Industrial Internet of Things. Moreover, few solutions provide the code on GitHub, while most do not provide any software availability information. In terms of privacy techniques used in these solutions, Table \ref{tab:privacy-techniques-works} summarizes that different encryption techniques have been employed to ensure privacy, including Public Key Encryption, Homomorphic Encryption, Proxy Re-Encryption, and Attribute-Based Encryption. Other techniques, such as anonymization, access control policies, and k-anonymity, are also used.

\subsubsection{\textbf{Healthcare and Personal Data}}

The analyzed works on healthcare and personal data privacy can be grouped into three main categories: decentralized personal data management, privacy preservation in healthcare data management systems, and secure data sharing and storage.

\begin{itemize}
  \item \textbf{Decentralized Personal Data Management:} The concept of using blockchain technology to protect privacy was first explored in 2015 by \citeauthor{2015-decentralizing-privacy} \cite{2015-decentralizing-privacy}, who proposed a decentralized personal data management system that utilizes a blockchain as an automated access-control manager to ensure that users own and control their data without relying on trusted third parties. Likewise, \citeauthor{10.1109/JIOT.2023.3242959} \cite{10.1109/JIOT.2023.3242959} introduced an anonymous authentication scheme leveraging the ElGamal commitment and the one-out-of-many proof to authorize data accessors while concealing their real identities from unauthorized entities and even the data owner. Employing a blockchain platform for data storage, access control lists, and storage addresses. Similarly, \citeauthor{10.1109/TR.2022.3190932} \cite{10.1109/TR.2022.3190932} proposed a personal data privacy protection scheme based on consortium blockchain that stores original data encrypted with an improved Paillier homomorphic encryption mechanism, namely PDPChain, where users realize fine-grained access control based on ciphertext policy attribute-based encryption (CP-ABE) on the blockchain.

  \item \textbf{Privacy Preservation in Healthcare Data Management Systems:} Some authors emphasized the sensitivity of healthcare data and the need to ensure privacy and security in handling this data. Works by \citeauthor{2016-healthcare-data-gateways} \cite{2016-healthcare-data-gateways}, \citeauthor{2017-Medibchain} \cite{2017-Medibchain}, and \citeauthor{2021-medical-data-sharing} \cite{2021-medical-data-sharing} all proposed models for preserving privacy in healthcare data management systems, with a focus on patient control and data ownership. They used access control mechanisms such as purpose-centric, patient-centered, and attribute-based approaches to share data securely.

\citeauthor{10.1109/TIFS.2023.3297327} \cite{10.1109/TIFS.2023.3297327}  proposed a designated-verifier aggregate signature scheme (named DVAS) based on permissioned blockchain to achieve sensitive data privacy. In this scheme, the aggregator can be used not only to aggregate signatures but also to sanitize data. Through smart contracts, the aggregator can sanitize the sensitive data according to the contract and convert the original signature of the sensitive data into a valid signature.
\citeauthor{10.1007/s12083-023-01521-w} \cite{10.1007/s12083-023-01521-w} proposed two schemes, keyword-based re-encryption (KRE) scheme and oblivious transfer-based re-encryption (OTRE) scheme with hybrid blockchain secure health data sharing. KRE preserves the data privacy of patients and provides keyword-based access with reduced key management overhead. OTRE is a superset of KRE that employs OT to preserve the privacy of users’ data and multiple data sharing
\citeauthor{10.1109/JIOT.2023.3287636} \cite{10.1109/JIOT.2023.3287636} proposed a searchable encryption scheme with multi-keyword search (MK-IPSE) based on blockchain to provide full privacy preservation and ciphertext retrieval for EMRs.  Some works have focused on secure data sharing and storage by combining cryptographic techniques with blockchain technology. Solutions for preserving privacy in personal health information (PHI) storage and sharing have been presented by \citeauthor{2018-towards-secure} \cite{2018-towards-secure} and \citeauthor{2020-covid-privacy} \cite{2020-covid-privacy}. The proposed systems by \citeauthor{2018-BPDS} \cite{2018-BPDS} and \citeauthor{2021-hOCBS} \cite{2021-hOCBS} combine cloud storage and blockchain technology for handling electronic medical records (EMRs) while preserving privacy. \citeauthor{2021-cp-bdhca} \cite{2021-cp-bdhca} introduced a solution that ensures secure access and analysis of electronic health records stored on healthcare cloud and application servers. Additionally, blockchain-based systems for sharing medical data, with a focus on privacy preservation, have been proposed by \citeauthor{2021-sp-chain} \cite{2021-sp-chain} and \citeauthor{2021-system-medical-data-privacy} \cite{2021-system-medical-data-privacy}. Finally, privacy-preserving frameworks for sharing electronic health records (EHRs), adopting encryption techniques and secure storage solutions, have been presented by \citeauthor{2021-medshare} \cite{2021-medshare} and \citeauthor{2022-scalable-bc-model-ipfs} \cite{2022-scalable-bc-model-ipfs}.

\end{itemize}

\subsubsection{\textbf{Internet of Things}}

The IoT, a network of devices equipped with sensors, software, and connectivity, enables the collection and sharing of data \cite{survey-IOT-ieee-com}. In this context, several authors have explored the potential of utilizing blockchain technology to enhance security, preserve privacy, and enable decentralization. However, developing a high-performing and scalable solution presents significant challenges as the number of devices and users grows. From the analyzed works, we grouped the approaches into the following subcategories based on their approaches and focus:

\begin{itemize}
  \item \textbf{Data Sharing and Access Control:} \citeauthor{2021-data-privacy-based-iot} \cite{2021-data-privacy-based-iot} and \citeauthor{2020-privy-sharing} \cite{2020-privy-sharing} both proposed frameworks for managing IoT devices and data sharing, with a focus on privacy preservation. They utilized smart contracts to validate connection rights and control access to user data. PrivySharing \cite{2020-privy-sharing} includes a reward system with digital tokens for users who share their data. Similarly, \citeauthor{10.1109/JIOT.2023.3307073}  \cite{10.1109/JIOT.2023.3307073} proposed Galaxy, a blockchain-based Pub/Sub IoT data sharing framework. To achieve Byzantine fault-tolerant (BFT) Pub/Sub, Galaxy adopts sharding to improve scalability and achieve efficient BFT Pub/Sub workflow within each shard with a novel leader rotation scheme.

  \item \textbf{Privacy-Preserving Techniques:} \citeauthor{2021-data-priv-iot} \cite{2021-data-priv-iot} proposed a solution for protecting IoT data privacy using proxy re-encryption and ring signature to protect the linkability and address information of the data-sharing parties. \citeauthor{2021-privacy-fog-iot} \cite{2021-privacy-fog-iot} suggested a distributed access control system based on blockchain technology that secures IoT data using fog computing, mixed linear and nonlinear spatiotemporal chaotic systems, and the least significant bit (LSB). \citeauthor{2022-bc-based-privacy-pres-reward} \cite{2022-bc-based-privacy-pres-reward} presented a privacy-preserving scheme and reward mechanism for sharing private data in IoT applications, where the data owner receives a reward for sharing data. \citeauthor{2021-priv-iot-homo} \cite{2021-priv-iot-homo} introduced a blockchain-based scheme incorporating data aggregation and homomorphic encryption in an IoT ecosystem. Similarly, \citeauthor{10.1109/JIOT.2023.3295763} \cite{10.1109/JIOT.2023.3295763}  presents a blockchain and federated learning-based data sharing scheme within a "cloud-edge-end" architecture to address terminal resource issues, introduces differential data sharing for resource-limited terminals. Similarly, \citeauthor{10.1109/JIOT.2021.3134755} \cite{10.1109/JIOT.2021.3134755} proposed a data-sharing privacy protection model (DS2PM) that is based on blockchain and a federated learning mechanism for solving privacy in data sharing.

  \item \textbf{Advanced Cryptographic Techniques:} \citeauthor{2020-privacy-remote-iot} \cite{2020-privacy-remote-iot} presented a scheme that leverages advanced cryptographic techniques, such as Lifted EC-ElGamal cryptosystem and bilinear pairing, in conjunction with blockchain technology to support efficient public batch signature verifications and protect IoT system security and data privacy.

  \item \textbf{Industrial IoT (IIoT) Environments:} Some authors focused on IoT devices in industrial environments. \citeauthor{2021-data-sec-sharing} \cite{2021-data-sec-sharing} and \citeauthor{2022-enterprise-ds-fabric-iot} \cite{2022-enterprise-ds-fabric-iot} proposed data security sharing models for the Industrial Internet of Things (IIoT) that incorporate privacy protection. They used hidden attribute authentication, off-chain databases, and encrypted storage solutions such as the Inter-Planetary File System (IPFS) to ensure secure data sharing and management.
\end{itemize}

\subsubsection{\textbf{Smart Grid}}

Some studies concentrated on addressing data privacy and security issues in the context of smart grids, with proposed solutions ranging from using advanced cryptographic techniques to implementing decentralized edge computing architectures. We can categorize the works into the following subcategories:

\begin{itemize}
    \item \textbf{Homomorphic Encryption-Based Solutions:}
    \citeauthor{2021-homomo-enc-smart-grid} \cite{2021-homomo-enc-smart-grid} proposed a deep learning and homomorphic encryption-based privacy-preserving data aggregation model for smart grids. The proposed model focuses on mitigating the negative impacts of a flash workload on the accuracy of prediction models and ensuring a secure data aggregation process with low computational overhead. Similarly, \citeauthor{10.1109/JIOT.2023.3247487} \cite{10.1109/JIOT.2023.3247487} proposed a cross-domain verifiable data collection and privacy-persevering sharing scheme based on lattice in blockchain-enhanced Smart Grids. The authors utilized lattice homomorphic encryption and spatial–temporal aggregation technology to enable cross-domain collection of data, dynamic addition, and deletion of users.

    \item \textbf{Edge Computing and Lightweight Schemes:}
    \citeauthor{2021-edge-light-smart-grid} \cite{2021-edge-light-smart-grid} presented an edge blockchain-assisted lightweight privacy-preserving data aggregation scheme for the smart grid EBDA, which combines homomorphic Paillier encryption and one-way hash chain techniques to improve data security and privacy protection in the edge layer of smart grids. This approach demonstrates the potential for decentralized edge computing architectures to enhance privacy and security in smart grid applications.

\end{itemize}

\subsubsection{\textbf{Alternative Applications}}

Other research has investigated the use of blockchain technology for various applications, such as cloud data provenance, unmanned aerial vehicle (UAV) data privacy, and government data sharing. We can group the works into the following subcategories based on their focus and approaches:

\begin{itemize}
    \item \textbf{Cloud Storage:}
    \citeauthor{2017-provchain} \cite{2017-provchain} introduced a decentralized and trusted cloud data provenance architecture called ProvChain, to provide tamper-proof records, enhance data accountability, and improve the availability of provenance data. Privacy is preserved through hashed user IDs, preventing blockchain network nodes from correlating data records with a specific user. Similarly, \citeauthor{10.1109/TSC.2022.3199111} \cite{10.1109/TSC.2022.3199111} introduces a blockchain framework on Ethereum to securely and privately query encrypted data on untrusted clouds.
    
    \item \textbf{Unmanned Aerial Vehicle (UAV):}
    \citeauthor{2021-drone-big-data} \cite{2021-drone-big-data} proposed a privacy protection scheme using blockchain technology and the NTHU-PCTRU (NTRU) cryptographic algorithm to protect the privacy of UAV big data. The scheme's security was validated through privacy analysis, and its performance was evaluated to show its effectiveness in key generation, encryption, and decryption. In addition, the authors demonstrated the security performance through the homomorphic property of the NTRU algorithm.
    
    \item \textbf{Government Data Sharing:}
    \citeauthor{2021-priv-gov-soc} \cite{2021-priv-gov-soc} proposed using a Service-On-Chain approach, which involves using smart contracts to establish data-sharing agreements between government departments. This approach aims to identify the data retrieval needs of different departments while ensuring the trustworthiness of the data and maintaining control over data ownership. This work highlights the potential for blockchain technology to improve privacy and efficiency in government data-sharing processes.

    \item \textbf{Transportation:} \citeauthor{10.1109/TITS.2022.3190487} \cite{10.1109/TITS.2022.3190487} combines zk-SNARKs, Merkle commitment scheme, and smart contract to facilitate privacy-preserving sharing of positioning data in maritime transportation systems.

    \item \textbf{Flight Operation:} \citeauthor{10.1109/JIOT.2023.3296460}  \cite{10.1109/JIOT.2023.3296460} presented a solution employing blockchain with zero-knowledge proof, proxy re-encryption, and other cryptographic methods for privacy protection in flight operation data.

    \item \textbf{Data Trading}: \citeauthor{10.1109/TC.2023.3251846} \cite{10.1109/TC.2023.3251846} introduced a blockchain-based data trading scheme where buyers specify data needs and sellers provide only the required data.

\end{itemize}

\begin{savenotes}
\begin{table*}[ht!]
\tiny
\centering
\caption{(Part I) Summary of blockchain-based privacy solutions sorted by publication year, including information on the use case, privacy technique employed, key contribution, blockchain platform used, and code availability.}

\label{tab:privacy-works}

\resizebox{1\textwidth}{!}{
    \begin{tabular}{lcclHHZ}
    \hline
    \textbf{Work} & \textbf{\begin{tabular}[c]{@{}c@{}}Publication\\ Year\end{tabular}} & \textbf{Use Case} & \textbf{Privacy Technique} & \textbf{Key Contribution} & \textbf{\begin{tabular}[c]{@{}c@{}}Blockchain platform\\ used by the authors\end{tabular}} & \textbf{\begin{tabular}[c]{@{}c@{}}Software\\ Availability\end{tabular}} \\ \hline
    \citeauthor{2015-decentralizing-privacy} \cite{2015-decentralizing-privacy} & 2015 & Personal Data & \begin{tabular}[c]{@{}l@{}}Data Encryption, Off-chain Storage \\ and Access Control \\\end{tabular}  & \begin{tabular}[c]{@{}l@{}}A decentralized personal data management\\  system with access-control\end{tabular} & N/A & N/A \\ \hline
    \citeauthor{2016-healthcare-data-gateways} \cite{2016-healthcare-data-gateways} & 2016 & Healthcare & \begin{tabular}[c]{@{}l@{}}Secure Multi-Party Computation (MPC) \\ Techniques \\\end{tabular} & \begin{tabular}[c]{@{}l@{}}A blockchain-based model to allow patients\\ to own and control their healthcare data\end{tabular} & N/A & N/A \\ \hline
    \citeauthor{2017-provchain} \cite{2017-provchain} & 2017 & Cloud Storage & Hashed user ID (Anonymization) & \begin{tabular}[c]{@{}l@{}}A blockchain-based system for securing and \\ auditing cloud data provenance\end{tabular} & N/A & N/A \\ \hline
    \citeauthor{2017-Medibchain} \cite{2017-Medibchain} & 2017 & Healthcare & \begin{tabular}[c]{@{}l@{}}Data Encryption and Access Control\\\end{tabular} & \begin{tabular}[c]{@{}l@{}}Patient-centered healthcare data management\\ system\end{tabular} & N/A & N/A \\ \hline
    \citeauthor{2018-BPDS} \cite{2018-BPDS} & 2018 & Healthcare & \begin{tabular}[c]{@{}l@{}}Public Key Encryption and\\ CP-ABE-based Access Control\end{tabular} & \begin{tabular}[c]{@{}l@{}}A system that utilizes both cloud storage and\\ blockchain technology to protect EMRs with \\ CP-ABE-based Access Control Mechanism\end{tabular} & N/A & N/A \\ \hline
    \citeauthor{2018-towards-secure} \cite{2018-towards-secure} & 2018 & Healthcare & \begin{tabular}[c]{@{}l@{}}Public Key Encryption with\\ Keyword Search (PEKS)\end{tabular} & \begin{tabular}[c]{@{}l@{}}A blockchain-based system for preserving the \\ privacy of personal health information (PHI)\end{tabular} & Ethereum & N/A \\ \hline
    \citeauthor{2020-privacy-remote-iot} \cite{2020-privacy-remote-iot} & 2020 & Internet of Things & \begin{tabular}[c]{@{}l@{}}Lifted EC-ElGamal Cryptosystem\\ and Bilinear pairing\end{tabular} & \begin{tabular}[c]{@{}l@{}}A scheme to protect the data privacy of the\\ IoT systems\end{tabular} & N/A & N/A \\ \hline
    \citeauthor{2020-privy-sharing} \cite{2020-privy-sharing} & 2020 & Internet of Things & \begin{tabular}[c]{@{}l@{}}Data Encryption and Hyperledger\\ Fabric Channels\end{tabular} & \begin{tabular}[c]{@{}l@{}}Framework to share IoT data with privacy \\ preservation through the use of channels and\\ smart contracts\end{tabular} & \begin{tabular}[c]{@{}c@{}}Hyperledger \\ Fabric\end{tabular} & N/A \\ \hline
    \citeauthor{2021-edge-light-smart-grid} \cite{2021-edge-light-smart-grid} & 2021 & Smart Grid & Homomorphic Encryption & \begin{tabular}[c]{@{}l@{}}Privacy-preserving data aggregation scheme for\\ use in smart grid\end{tabular} & N/A & N/A \\ \hline
    \citeauthor{2021-privacy-fog-iot} \cite{2021-privacy-fog-iot} & 2021 & Internet of Things & Access control Policies and Encryption & \begin{tabular}[c]{@{}l@{}}Framework for secure access control of IoT data\\ using blockchain technology\end{tabular} & \begin{tabular}[c]{@{}c@{}}Hyperledger \\ Fabric\end{tabular} & N/A \\ \hline
    \citeauthor{2021-data-priv-iot} \cite{2021-data-priv-iot} & 2021 & Internet of Things & Anonymization and Data Encryption & \begin{tabular}[c]{@{}l@{}}Privacy protection scheme for the IoT using\\ encryption\end{tabular} & N/A & N/A \\ \hline
    \citeauthor{2021-data-sec-sharing} \cite{2021-data-sec-sharing} & 2021 & \begin{tabular}[c]{@{}c@{}}Industrial Internet of \\ Things\end{tabular} & \begin{tabular}[c]{@{}l@{}}Encryption and Hidden Attribute\\ Authentication\end{tabular} & Privacy protection model for IIoT applications & N/A & N/A \\ \hline
    \citeauthor{2021-drone-big-data} \cite{2021-drone-big-data} & 2021 & UAV big data & \begin{tabular}[c]{@{}l@{}}Encryption, Threshold Scheme, \\ NTRU Cryptosystem,\end{tabular} & Privacy protection scheme for UAV big data & N/A & N/A \\ \hline
    \citeauthor{2021-homomo-enc-smart-grid} \cite{2021-homomo-enc-smart-grid} & 2021 & Smart Grid & Homomorphic Encryption & \begin{tabular}[c]{@{}l@{}}Homomorphic encryption-based data aggregation\\ (BHDA) scheme\end{tabular} & N/A & N/A \\ \hline
    \citeauthor{2021-hOCBS} \cite{2021-hOCBS} & 2021 & Healthcare & \begin{tabular}[c]{@{}l@{}}Preserving Anonymity and Dynamic\\ Consent Management\end{tabular} & \begin{tabular}[c]{@{}l@{}}Hybrid privacy-preserving framework for off-chain\\ and on-chain systems (hOCBS) for healthcare data \\ management\end{tabular} & N/A & N/A \\ \hline
    \citeauthor{2021-sp-chain} \cite{2021-sp-chain} & 2021 & Healthcare & Proxy Re-encryption Scheme & \begin{tabular}[c]{@{}l@{}}A blockchain-based system to enable patients and\\ doctors to share data\end{tabular} & N/A & N/A \\ \hline
    \citeauthor{2021-priv-gov-soc} \cite{2021-priv-gov-soc} & 2021 & Government & \begin{tabular}[c]{@{}l@{}}Smart Contracts to Validate\\ Permission Settings\end{tabular} & \begin{tabular}[c]{@{}l@{}}Government data sharing in a consortium blockchain\\ system\end{tabular} & \begin{tabular}[c]{@{}c@{}}Hyperledger \\ Fabric\end{tabular} & N/A \\ \hline

    \end{tabular}
}
\end{table*}
\end{savenotes}

\begin{savenotes}
\begin{table*}[ht!]
\tiny
\centering
\caption{(Part II) Summary of blockchain-based privacy solutions sorted by publication year, including information on the use case, privacy technique employed, key contribution, blockchain platform used, and code availability.}

 \setlength{\tabcolsep}{1.2em}
\def\arraystretch{1.2}
\label{tab:privacy-works-2}

\resizebox{1\textwidth}{!}{
    \begin{tabular}{lcclHHZ}
    \hline
    \textbf{Work} & \textbf{\begin{tabular}[c]{@{}c@{}}Publication\\ Year\end{tabular}} & \textbf{Use Case} & \textbf{Privacy Technique} & \textbf{Key Contribution} & \textbf{\begin{tabular}[c]{@{}c@{}}Blockchain platform\\ used by the authors\end{tabular}} & \textbf{\begin{tabular}[c]{@{}c@{}}Software\\ Availability\end{tabular}} \\ \hline

        \citeauthor{2021-data-privacy-based-iot} \cite{2021-data-privacy-based-iot} & 2021 & Internet of Things & \begin{tabular}[c]{@{}l@{}}Smart Contracts to Validate\\ Permission Settings\end{tabular} & \begin{tabular}[c]{@{}l@{}}Privacy-Preserving Framework for IoT device\\ management using blockchain technology\end{tabular} & Ethereum & N/A \\ \hline
    \citeauthor{2021-medshare}\footnote{\url{https://github.com/groupjia/medshare2021}} \cite{2021-medshare} & 2021 & Healthcare & Attribute-based Encryption Scheme & \begin{tabular}[c]{@{}l@{}}Design and implementation of a multi-keyword boolean\\ search operation over encrypted EHRs\end{tabular} & Ethereum & \textit{\begin{tabular}[c]{@{}c@{}}https://github.com/groupjia/\\ medshare2021\end{tabular}} \\ \hline
    \citeauthor{2021-priv-iot-homo}\footnote{\url{https://github.com/Floukil/E2EAggregation}} \cite{2021-priv-iot-homo} & 2021 & Internet of Things & Homomorphic Encryption & \begin{tabular}[c]{@{}l@{}}Homomorphic encryption-based privacy-preserving\\ IoT data aggregation scheme\end{tabular} & Ethereum & \textit{\begin{tabular}[c]{@{}c@{}}https://github.com/Floukil/\\ E2EAggregation\end{tabular}} \\ \hline
    \citeauthor{2021-system-medical-data-privacy}\footnote{\url{https://github.com/nanodaemony/medicalledger}} \cite{2021-system-medical-data-privacy} & 2021 & Healthcare & Proxy Re-encryption Scheme & \begin{tabular}[c]{@{}l@{}}A blockchain-based medical data information system\\ for secure collection, storage, query, and sharing\end{tabular} & \begin{tabular}[c]{@{}c@{}}Hyperledger \\ Fabric\end{tabular} & \textit{\begin{tabular}[c]{@{}c@{}}https://github.com/nanodaemony/\\ medicalledger\end{tabular}} \\ \hline
    \citeauthor{2021-medical-data-sharing}\footnote{\url{https://github.com/mythsand/privacy-preserving-medical-data}} \cite{2021-medical-data-sharing} & 2021 & Healthcare & K-anonymity and Searchable Encryption & \begin{tabular}[c]{@{}l@{}}Privacy-preserving medical data sharing scheme\\ using K-anonymity, searchable encryption, and ABAC\end{tabular} & \begin{tabular}[c]{@{}c@{}}Hyperledger \\ Fabric\end{tabular} & \textit{\begin{tabular}[c]{@{}c@{}}https://github.com/mythsand/\\ privacy-preserving-medical-data\end{tabular}} \\ \hline
    \citeauthor{2020-covid-privacy} \cite{2020-covid-privacy} & 2022 & Healthcare & Attribute-based encryption scheme & \begin{tabular}[c]{@{}l@{}}Proposed a scheme for secure and private storage\\ and sharing of COVID-19 EMRs\end{tabular} & Ethereum & N/A \\ \hline
    \citeauthor{2022-enterprise-ds-fabric-iot} \cite{2022-enterprise-ds-fabric-iot} & 2022 & \begin{tabular}[c]{@{}c@{}}Industrial Internet of \\ Things\end{tabular} & \begin{tabular}[c]{@{}l@{}}Data Encryption and Hyperledger\\ Fabric Channels\end{tabular} & Scheme to protect privacy in IIOT applications & \begin{tabular}[c]{@{}c@{}}Hyperledger \\ Fabric\end{tabular} & N/A \\ \hline
    \citeauthor{2022-scalable-bc-model-ipfs} \cite{2022-scalable-bc-model-ipfs} & 2022 & Healthcare & Data Encryption and Patient Permission & Framework for secure EHR management in healthcare & N/A & N/A \\ \hline
    \citeauthor{2021-cp-bdhca} \cite{2021-cp-bdhca} & 2022 & Healthcare & \begin{tabular}[c]{@{}l@{}}Elliptic Curve Cryptographic and\\ Two-step Authentication\end{tabular} & \begin{tabular}[c]{@{}l@{}}Privacy-preserving scheme combining blockchain,\\ ECC, and 2FA\end{tabular} & N/A & N/A \\ \hline
    \citeauthor{2022-bc-based-privacy-pres-reward} \cite{2022-bc-based-privacy-pres-reward} & 2022 & Internet of Things & Ring Signatures and Monero & \begin{tabular}[c]{@{}l@{}}Scheme for IoT with anonymity, unforgeability,\\ non-frameability, and behavior prevention\end{tabular} & \begin{tabular}[c]{@{}c@{}}Hyperledger \\ Fabric\end{tabular} & N/A \\ \hline

    \citeauthor{10.1109/TR.2022.3190932} \cite{10.1109/TR.2022.3190932} & 2022 & Personal Data & Homomorphic Encryption \\ \hline
    \citeauthor{10.1109/JIOT.2023.3242959} \cite{10.1109/JIOT.2023.3242959} & 2023 & Personal Data & ElGamal Commitment \\ \hline
    \citeauthor{10.1109/TIFS.2023.3297327} \cite{10.1109/TIFS.2023.3297327} & 2023 & Healthcare & Aggregate Signatures \\ \hline
    \citeauthor{10.1007/s12083-023-01521-w} \cite{10.1007/s12083-023-01521-w} & 2023 & Healthcare & \begin{tabular}[c]{@{}l@{}}keyword-based re-encryption \\ and oblivious transfer-based\\re-encryption\end{tabular} \\ \hline

    \citeauthor{10.1109/JIOT.2023.3287636} \cite{10.1109/JIOT.2023.3287636} & 2023 & Healthcare & Searchable Encryption Scheme \\ \hline

    \citeauthor{10.1109/JIOT.2023.3307073} \cite{10.1109/JIOT.2023.3307073} & 2023 & Internet of Things & \begin{tabular}[c]{@{}l@{}}Threshold encryption and \\ access control list\end{tabular}
      \\ \hline

    \citeauthor{10.1109/JIOT.2023.3295763} \cite{10.1109/JIOT.2023.3295763} & 2023 & Internet of Things & Federated Learning  \\ \hline

    \citeauthor{10.1109/JIOT.2021.3134755} \cite{10.1109/JIOT.2021.3134755} & 2023 & Internet of Things & Federated Learning  \\ \hline

    \citeauthor{10.1109/JIOT.2023.3247487} \cite{10.1109/JIOT.2023.3247487} & 2023 & Smart Grids & Homomorphic Encryption  \\ \hline
    
    \citeauthor{10.1109/TSC.2022.3199111} \cite{10.1109/TSC.2022.3199111} & 2023 & Cloud Outsourcing & \begin{tabular}[c]{@{}l@{}}Encrypted Index Storage and \\ Stealth Authorization Scheme\end{tabular}  \\ \hline

    \citeauthor{10.1109/TITS.2022.3190487} \cite{10.1109/TITS.2022.3190487} & 2023 & Maritime Transportation  &  \begin{tabular}[c]{@{}l@{}}zk-SNARK and \\ Commitment-related technologies\end{tabular}  \\ \hline

    \citeauthor{10.1109/JIOT.2023.3296460}  \cite{10.1109/JIOT.2023.3296460} & 2023 & Flight Operation  & \begin{tabular}[c]{@{}l@{}}Hash Anonymous Identity, \\ zk-SNARKs and Proxy Re-Encryption\end{tabular}  \\ \hline

    \citeauthor{10.1109/TC.2023.3251846} \cite{10.1109/TC.2023.3251846} & 2023 & Data Trading  & \begin{tabular}[c]{@{}l@{}}Attribute-Based Credentials  Encryption\end{tabular}\\ \hline

    \end{tabular}
}
\end{table*}
\end{savenotes}

\begin{savenotes}
\begin{table*}[ht!]
\centering
\caption{Grouped Summary of Blockchain-Based Privacy Solutions by Privacy Technique}
\label{tab:privacy-techniques-works}

    \begin{tabular}{ll}
    \hline
    \textbf{Privacy Technique} & \textbf{Works Employing the Technique} \\ \hline
    Data Encryption & \cite{2015-decentralizing-privacy}, \cite{2017-Medibchain}, \cite{2020-privacy-remote-iot}, \cite{2021-data-priv-iot}, \cite{2022-scalable-bc-model-ipfs}, \cite{2022-enterprise-ds-fabric-iot}, \cite{2020-privy-sharing} \\ \hline
    Homomorphic Encryption & \cite{10.1109/TR.2022.3190932}, \cite{2021-edge-light-smart-grid}, \cite{2021-homomo-enc-smart-grid}, \cite{10.1109/JIOT.2023.3247487}, \cite{2021-priv-iot-homo} \\ \hline
    Access Control & \cite{2015-decentralizing-privacy}, \cite{2017-Medibchain}, \cite{2021-privacy-fog-iot}, \cite{2021-priv-gov-soc}, \cite{2021-data-privacy-based-iot} \\ \hline
    Secure Multi-Party Computation (MPC) & \cite{2016-healthcare-data-gateways} \\ \hline
    Public Key Encryption & \cite{2018-BPDS}, \cite{2018-towards-secure} \\ \hline
    Anonymization & \cite{2017-provchain}, \cite{2021-data-priv-iot} \\ \hline
    Federated Learning & \cite{10.1109/JIOT.2023.3295763}, \cite{10.1109/JIOT.2021.3134755} \\ \hline
    Attribute-based Encryption & \cite{2021-medshare}, \cite{2020-covid-privacy} \\ \hline
    zk-SNARKs and Related Technologies & \cite{10.1109/TITS.2022.3190487}, \cite{10.1109/JIOT.2023.3296460} \\ \hline
    Proxy Re-Encryption & \cite{2021-sp-chain}, \cite{2021-system-medical-data-privacy} \\ \hline
    Searchable Encryption & \cite{10.1109/JIOT.2023.3287636}, \cite{2021-medical-data-sharing} \\ \hline
    Elliptic Curve Cryptography & \cite{2021-cp-bdhca} \\ \hline
    Ring Signatures & \cite{2022-bc-based-privacy-pres-reward} \\ \hline
    K-anonymity & \cite{2021-medical-data-sharing} \\ \hline
    ElGamal Commitment & \cite{10.1109/JIOT.2023.3242959} \\ \hline
    Aggregate Signatures & \cite{10.1109/TIFS.2023.3297327} \\ \hline
    Oblivious Transfer-based Re-encryption & \cite{10.1007/s12083-023-01521-w} \\ \hline
    Threshold Encryption & \cite{10.1109/JIOT.2023.3307073} \\ \hline
    Encrypted Index Storage & \cite{10.1109/TSC.2022.3199111} \\ \hline
    Stealth Authorization Scheme & \cite{10.1109/TSC.2022.3199111} \\ \hline
    Attribute-Based Credentials Encryption & \cite{10.1109/TC.2023.3251846} \\ \hline
    Smart Contracts & \cite{2021-priv-gov-soc}, \cite{2021-data-privacy-based-iot} \\ \hline
    CP-ABE-based Access Control & \cite{2018-BPDS} \\ \hline
    Keyword-based Re-encryption & \cite{10.1007/s12083-023-01521-w} \\ \hline
    Hash Anonymous Identity & \cite{10.1109/JIOT.2023.3296460} \\ \hline
    \end{tabular}

\end{table*}
\end{savenotes}

\subsection{Consent Management in Blockchain-based and Privacy-sensitive Applications}
\label{sec:consent}
Data sharing between various institutions has become increasingly common with the recent proliferation of digital technologies. This section explores the mechanisms to enable permission for data sharing among stakeholders in different use cases. Specifically, we will focus on how blockchain technology manages user consent in these contexts. Tables \ref{tab:consent-works} and \ref{tab:consent-works-2} summarize the consent management works, including details such as the year of publication, the specific use case studied, the main contributions, the blockchain platform used for prototyping, and whether the code is publicly available. Most of the works focus on the healthcare sector, with several papers exploring academic records, IoT, and data sharing in general. Other domains include financial services and open banking. 

\subsubsection{\textbf{Healthcare}}

A number of studies in the literature have investigated the potential of blockchain technology to enhance patient consent management in the context of data sharing, particularly in the healthcare industry. From the analysis, these studies can be grouped into three main categories: electronic medical records (EMR) and personal health record (PHR) management, clinical trials and research, biobanking, and genomic data sharing.

\begin{itemize}
    \item \textbf{EMR and PHR Management:} Some works have focused on decentralized architectures for electronic medical records management (EMR) and personal health record (PHR) management, ensuring privacy and compliance with data protection regulations. For example, \citeauthor{2016-medrec} \cite{2016-medrec} proposed MedRec, while \citeauthor{2019-medchain} \cite{2019-medchain} proposed a system utilizing ElGamal re-encryption. Other notable solutions include the e-Health consent management framework by \citeauthor{2020-consent-design-bc-based-consent-manag} \cite{2020-consent-design-bc-based-consent-manag} and the individual consent model by \citeauthor{2020-consent-health} \cite{2020-consent-health}. These works aim to provide patients with control over data sharing and ensure compliance with relevant regulations, such as HIPAA and GDPR. Similarly, \citeauthor{10.1109/ICOIN56518.2023.10048920} \cite{10.1109/ICOIN56518.2023.10048920} proposed a blockchain-based framework for consent management called Heimdall. 

    \item \textbf{Clinical Trials and Research:} Some works have explored the potential of blockchain technology for secure data sharing and analysis in medical research, automating aspects of consent and data management. Some examples include the PHR platform by \citeauthor{10.3837/TIIS.2021.12.008} \cite{10.3837/TIIS.2021.12.008}, ConsentChain by \citeauthor{2021-dynamic-consent-genomic} \cite{2021-dynamic-consent-genomic}, and DynamiChain by \citeauthor{2021-dynamichain} \cite{2021-dynamichain}. Additionally, \citeauthor{10.1016/j.hcc.2022.100084} \cite{10.1016/j.hcc.2022.100084} proposed a multi-hop permission delegation scheme with the controllable delegation for electronic health record (EHR) sharing.

    \item \textbf{Biobanking and Genomic Data Sharing:} A few studies have addressed the specific challenges of managing patient consent in biobanking and genomic data sharing. For example, \citeauthor{10.3389/fmed.2022.837197} \cite{10.3389/fmed.2022.837197} proposed the dynamic consent platform METORY, while \citeauthor{2022-crowmedII-consent} \cite{2022-crowmedII-consent} proposed a framework called CrowMed-II, which explores patient consent to share data using smart contracts. Furthermore, \citeauthor{2020-consent-biobanking} \cite{2020-consent-biobanking} proposed Dwarna, a dynamic consent architecture that enables connections between various stakeholders of the Malta Biobank while adhering to GDPR.

\end{itemize}

All these studies share a similar focus on developing secure and patient-centric solutions for data management in the healthcare industry through blockchain technology, smart contracts, and encryption techniques. They also share a common goal of providing control over personal data to patients by enabling them to share their data with certain parties as desired.

\subsubsection{\textbf{Internet of Things and Other Applications}}

Some authors have examined its features in the context of consent management and data privacy across various domains, such as open banking, online social networks, secure data sharing, fitness apps, financial services, data trading, education, and active assisted living.

\begin{itemize}
    \item \textbf{Open Banking and Online Social Networks:} \citeauthor{10.1007/978-981-15-9317-8_3} \cite{10.1007/978-981-15-9317-8_3} proposed a framework to monitor customer attributes and ensure transparency in open banking operations. Similarly, \citeauthor{10.1109/ICICT50521.2020.00054} \cite{10.1109/ICICT50521.2020.00054} conducted a study that compared the standards for valid consent outlined in the GDPR with the consent-gathering practices currently employed by online social networks (OSNs) and explored ways in which blockchain technology could be utilized to ensure compliance with GDPR.
    
    \item \textbf{Secure Data Sharing and Permission Management:} \cite{2018-double-blind}, \cite{10.1109/INISTA.2018.8466268}, \cite{2018-advocate}, and \cite{2019-blockchain-permission-iot},  focus on enabling customers to grant or revoke access to their data using smart contracts and consent-driven, double-blind data sharing on platforms like Hyperledger Fabric.
    
    \item \textbf{Fitness Apps, Financial Services, and Data Trading:}
    Some studies address existing privacy policies, consent practices, and the need for automated and proactive revocation of permissions in industries like fitness apps, financial services, and data trading \cite{10.1145/3511616.3513100}, \cite{10.1109/SMDS53860.2021.00015}, \cite{10.1109/COMSNETS51098.2021.9352817}, \cite{10.1007/978-3-030-78621-2_54}, and \cite{10.1109/QRS-C55045.2021.00159}
    
    \item \textbf{Education and Active Assisted Living:}
    Other studies propose solutions for consent management in education and active assisted living (AAL) contexts. \citeauthor{10.1109/ICCCI.2018.8441445} \cite{10.1109/ICCCI.2018.8441445} proposed a framework using Hyperledger Fabric and smart contracts to grant authorization rights for student data privacy. \citeauthor{2019-unichain-blockchain-acad-records} proposed UniChain \cite{2019-unichain-blockchain-acad-records}, a blockchain-based system for managing academic records with permission management. Additionally, \citeauthor{2020-blockchain-assisted-living} \cite{2020-blockchain-assisted-living} presented a conceptual framework addressing trust issues in the consent management process of AAL technologies, utilizing blockchain technology to ensure the secure storage and transmission of data.

    \item \textbf{Human-centered:} \citeauthor{10.1145/3576842.3582379} \cite{10.1145/3576842.3582379} presented a privacy-preserving model that leverages the blockchain features for consent management and transparency in Human-Centered Internet of Things environments.

\end{itemize}

\citeauthor{10.1016/j.jlamp.2023.100886} \cite{10.1016/j.jlamp.2023.100886} proposed a privacy-enabled formal data-sharing model for the consent management system. Additionally, the authors presented a middleware implemented using Solidity and Python REST API with GDPR compliance.

\begin{savenotes}
\begin{table*}[h!]
\tiny
\centering
\caption{(Part I) Summary of blockchain-based consent management works sorted by publication year, including information on the use case, key contribution, blockchain platform used, and software availability}
\setlength{\tabcolsep}{1.5em}
\def\arraystretch{1.1}
\label{tab:consent-works}

\resizebox{\textwidth}{!}{

\begin{tabular}{lcclHZ}
\hline
\multicolumn{1}{c}{\textbf{Work}}                                                                         & \textbf{\begin{tabular}[c]{@{}c@{}}Publication\\ Year\end{tabular}} & \textbf{Use Case}                                                      & \textbf{Key Contribution}                                                                                                                                                        & \textbf{\begin{tabular}[c]{@{}c@{}}Blockchain platform\\ used by the authors\end{tabular}} & \textbf{\begin{tabular}[c]{@{}c@{}}Software\\ Availability\end{tabular}}                             \\ \hline
\citeauthor{2016-medrec} \cite{2016-medrec}                                                               & 2016                                                                & Healthcare                                                             & Decentralized EMR with patient control                                                                                                                                           & Ethereum                                                               & N/A                                                                                                  \\ \hline
\citeauthor{10.1109/ICCCI.2018.8441445} \cite{10.1109/ICCCI.2018.8441445}                                 & 2018                                                                & \begin{tabular}[c]{@{}c@{}}Academic\\ Records\end{tabular}             & \begin{tabular}[c]{@{}l@{}}Framework for nested authorization, enabling \\ a school administrator to grant data access \\ rights to third party programs\end{tabular}            & \begin{tabular}[c]{@{}c@{}}Hyperledger\\ Fabric\end{tabular}           & N/A                                                                                                  \\ \hline
\citeauthor{2017-clinical-trials}\footnote{\url{https://github.com/benchoufi/DocChain}} \cite{2017-clinical-trials}                                             & 2018                                                                & \begin{tabular}[c]{@{}c@{}}Healthcare\\ (Clinical Trials)\end{tabular} & \begin{tabular}[c]{@{}l@{}}Proof-of-concept protocol for consent collection\\ in clinical trials\end{tabular}                                                                    & N/A                                                                    & \textit{\begin{tabular}[c]{@{}c@{}}https://github.com/benchoufi/\\ DocChain\end{tabular}}            \\ \hline
\citeauthor{10.1109/INISTA.2018.8466268} \cite{10.1109/INISTA.2018.8466268}                               & 2018                                                                & Internet of Things                                                                    & \begin{tabular}[c]{@{}l@{}}Design and implementation of a Forms of Consent \\ application, that interacts with a set of Smart Contracts\end{tabular}                             & Ethereum                                                               & N/A                                                                                                  \\ \hline
\citeauthor{2018-double-blind} \cite{2018-double-blind}                                                   & 2018                                                                & Know-your-customer                                                                    & \begin{tabular}[c]{@{}l@{}}Consent-driven and double-blind data sharing in\\ corporate KYC use case\end{tabular}                                                                 & \begin{tabular}[c]{@{}c@{}}Hyperledger\\ Fabric\end{tabular}           & N/A                                                                                                  \\ \hline
\citeauthor{2019-unichain-blockchain-acad-records} \cite{2019-unichain-blockchain-acad-records}           & 2019                                                                & \begin{tabular}[c]{@{}c@{}}Academic\\ Records\end{tabular}             & \begin{tabular}[c]{@{}l@{}}Blockchain-based System for Electronic\\ Academic Records\end{tabular}                                                                                & Ethereum                                                               & N/A                                                                                                  \\ \hline
\citeauthor{2019-medchain} \cite{2019-medchain}                                                           & 2019                                                                & Healthcare                                                             & \begin{tabular}[c]{@{}l@{}}System to manage medical records, incentive \\ mechanisms and access control\end{tabular}                                                              & Ethereum                                                               & N/A                                                                                                  \\ \hline
\citeauthor{2018-advocate}\footnote{\url{https://github.com/AnthonyK95/adplatform}} \cite{2018-advocate}                                                           & 2019                                                                & Internet of Things                                                                     & User-centric with GDPR requirements                                                                                                                                              & Ethereum                                                               & \textit{\begin{tabular}[c]{@{}c@{}}https://github.com/AnthonyK95/\\ adplatform\end{tabular}}         \\ \hline
\citeauthor{2019-blockchain-permission-iot} \cite{2019-blockchain-permission-iot}                         & 2019                                                                & Internet of Things                                                                     & \begin{tabular}[c]{@{}l@{}}Decentralized architecture for permission delegation\\ and access control for IoT applications\end{tabular}                                           & \begin{tabular}[c]{@{}c@{}}Hyperledger\\ Fabric\end{tabular}           & N/A                                                                                                  \\ \hline
\citeauthor{2020-blockchain-assisted-living} \cite{2020-blockchain-assisted-living}                       & 2020                                                                & \begin{tabular}[c]{@{}c@{}}Active Assisted\\ Living (AAL)\end{tabular} & \begin{tabular}[c]{@{}l@{}}A conceptual framework for consent management \\ active assisted living (AAL) technologies\end{tabular}                                               & \begin{tabular}[c]{@{}c@{}}Hyperledger\\ Fabric\end{tabular}           & N/A                                                                                                  \\ \hline
\citeauthor{2020-consent-design-bc-based-consent-manag} \cite{2020-consent-design-bc-based-consent-manag} & 2020                                                                & Healthcare                                                             & \begin{tabular}[c]{@{}l@{}}Framework for managing patient consent that is \\ compliant with privacy regulations\end{tabular}                                                     & \begin{tabular}[c]{@{}c@{}}Hyperledger\\ Fabric\end{tabular}           & N/A                                                                                                  \\ \hline
\citeauthor{consent-health-data-sharing}\footnote{\url{https://gitlab.com/vjaiman/consentblockchainluce}} \cite{consent-health-data-sharing}                               & 2020                                                                & Healthcare                                                             & \begin{tabular}[c]{@{}l@{}}Architecture combining two consent representation\\ models DUO and ADA-M\end{tabular}                                                                 & Ethereum                                                               & \textit{\begin{tabular}[c]{@{}c@{}}https://gitlab.com/vjaiman/\\ consentblockchainluce\end{tabular}} \\ \hline
\citeauthor{2020-patient-consent-e-health} \cite{2020-patient-consent-e-health}                           & 2020                                                                & Healthcare                                                             & \begin{tabular}[c]{@{}l@{}}Prototype to allow patient to manage consent for\\ other institutions\end{tabular}                                                                    & \begin{tabular}[c]{@{}c@{}}Hyperledger\\ Fabric\end{tabular}           & N/A                                                                                                  \\ \hline
\citeauthor{2020-consent-biobanking}\footnote{\url{https://github.com/NicholasMamo/dwarna}} \cite{2020-consent-biobanking}                                       & 2020                                                                & Healthcare                                                             & \begin{tabular}[c]{@{}l@{}}Solution for researchers consent to share\\ biomedical studies\end{tabular}                                                                           & \begin{tabular}[c]{@{}c@{}}Hyperledger\\ Fabric\end{tabular}           & \textit{\begin{tabular}[c]{@{}c@{}}https://github.com/NicholasMamo/\\ dwarna\end{tabular}}           \\ \hline
\citeauthor{2020-consent-clinical-trials} \cite{2020-consent-clinical-trials}                             & 2020                                                                & \begin{tabular}[c]{@{}c@{}}Healthcare\\ (Clinical Trials)\end{tabular} & Dynamic consent management in clinical trials                                                                                                                                    & \begin{tabular}[c]{@{}c@{}}Hyperledger\\ Fabric\end{tabular}           & N/A                                                                                                  \\ \hline
\citeauthor{10.1109/ICICT50521.2020.00054} \cite{10.1109/ICICT50521.2020.00054}                           & 2020                                                                & \begin{tabular}[c]{@{}c@{}}Online Social\\ Networks\end{tabular}       & \begin{tabular}[c]{@{}l@{}}A comparative study of GDPR and OSN consent\\ practices\end{tabular}                                                                                  & N/A                                                                    & N/A                                                                                                  \\ \hline
\end{tabular}
}
\end{table*}
\end{savenotes}

\begin{savenotes}
\begin{table*}[h!]
\tiny
\centering
\caption{(Part II) Summary of blockchain-based consent management works sorted by publication year, including information on the use case, key contribution, blockchain platform used, and software availability}
\setlength{\tabcolsep}{1.5em}
\def\arraystretch{1.1}
\label{tab:consent-works-2}

\resizebox{\textwidth}{!}{

\begin{tabular}{lcclHZ}
\hline
\multicolumn{1}{c}{\textbf{Work}}                                                                         & \textbf{\begin{tabular}[c]{@{}c@{}}Publication\\ Year\end{tabular}} & \textbf{Use Case}                                                      & \textbf{Key Contribution}                                                                                                                                                        & \textbf{\begin{tabular}[c]{@{}c@{}}Blockchain platform\\ used by the authors\end{tabular}} & \textbf{\begin{tabular}[c]{@{}c@{}}Software\\ Availability\end{tabular}}                             \\ \hline
\citeauthor{10.1109/COMSNETS51098.2021.9352817} \cite{10.1109/COMSNETS51098.2021.9352817}                 & 2021                                                                & \begin{tabular}[c]{@{}c@{}}Data Sharing\\ in General\end{tabular}      & \begin{tabular}[c]{@{}l@{}}End-to-end protocol for secure, fair data trading\\ using blockchain technology addressing consent,\\ traceability, and fairness.\end{tabular}        & Ethereum                                                               & N/A                                                                                                  \\ \hline
\citeauthor{10.1007/978-3-030-78621-2_54}  \cite{10.1007/978-3-030-78621-2_54}                            & 2021                                                                & \begin{tabular}[c]{@{}c@{}}Data Sharing\\ in General\end{tabular}      & \begin{tabular}[c]{@{}l@{}}On-chain user permission management mechanism \\ for blockchain with automatic and active \\ permission revocation using smart contracts\end{tabular} & N/A                                                                    & N/A                                                                                                  \\ \hline
\citeauthor{10.1109/QRS-C55045.2021.00159} \cite{10.1109/QRS-C55045.2021.00159}                           & 2021                                                                & \begin{tabular}[c]{@{}c@{}}Data Sharing\\ in General\end{tabular}      & User-friendly consent management model                                                                                                                                           & N/A                                                                    & \textit{\begin{tabular}[c]{@{}c@{}}https://github.com/crim-ca/\\ blockchain\end{tabular}}            \\ \hline
\citeauthor{10.1109/SMDS53860.2021.00015} \cite{10.1109/SMDS53860.2021.00015}                             & 2021                                                                & \begin{tabular}[c]{@{}c@{}}Financial\\ Services\end{tabular}           & \begin{tabular}[c]{@{}l@{}}Automated consent management model for\\ financial services\end{tabular}                                                                              & N/A                                                                    & N/A                                                                                                  \\ \hline
\citeauthor{10.1136/medethics-2019-105963} \cite{10.1136/medethics-2019-105963}                           & 2021                                                                & Healthcare                                                             & \begin{tabular}[c]{@{}l@{}}``Prosent" concept for pseudonymous, proactive\\ data consent in medical research using\\ blockchain technology\end{tabular}                           & N/A                                                                    & N/A                                                                                                  \\ \hline
\citeauthor{10.3837/TIIS.2021.12.008} \cite{10.3837/TIIS.2021.12.008}                                     & 2021                                                                & Healthcare                                                             & \begin{tabular}[c]{@{}l@{}}Blockchain-based PHR platform designed for\\ patient control of personal information\end{tabular}                                                     & \begin{tabular}[c]{@{}c@{}}Hyperledger\\ Fabric\end{tabular}           & N/A                                                                                                  \\ \hline
\citeauthor{2021-dynamic-consent-genomic}\footnote{\url{https://data.mendeley.com/datasets/vwy3hj5h8n/1}} \cite{2021-dynamic-consent-genomic}                             & 2021                                                                & Healthcare                                                             & \begin{tabular}[c]{@{}l@{}}Blockchain-based clinical genomic data sharing\\ architecture\end{tabular}                                                                            & Ethereum                                                               & \textit{\begin{tabular}[c]{@{}c@{}}https://data.mendeley.com/datasets/\\ vwy3hj5h8n/1\end{tabular}}  \\ \hline
\citeauthor{2021-dynamichain} \cite{2021-dynamichain}                                                     & 2021                                                                & Healthcare                                                             & \begin{tabular}[c]{@{}l@{}}Dynamic consent managment for handling health\\ examination data\end{tabular}                                                                         & \begin{tabular}[c]{@{}c@{}}Hyperledger\\ Fabric\end{tabular}           & N/A                                                                                                  \\ \hline
\citeauthor{10.1007/978-981-15-9317-8_3} \cite{10.1007/978-981-15-9317-8_3}                               & 2021                                                                & \begin{tabular}[c]{@{}c@{}}Open\\ Banking\end{tabular}                 & \begin{tabular}[c]{@{}l@{}}Framework for customer consent in open\\ banking process\end{tabular}                                                                                 & \begin{tabular}[c]{@{}c@{}}Hyperledger\\ Fabric\end{tabular}           & N/A                                                                                                  \\ \hline
\citeauthor{10.1145/3511616.3513100} \cite{10.1145/3511616.3513100}                                       & 2022                                                                & \begin{tabular}[c]{@{}c@{}}Fitness\\ Ecosystem\end{tabular}            & \begin{tabular}[c]{@{}l@{}}Problems Identification in current privacy policies\\ and consent practices in fitness apps\end{tabular}                                              & N/A                                                                    & N/A                                                                                                  \\ \hline
\citeauthor{10.1016/j.hcc.2022.100084} \cite{10.1016/j.hcc.2022.100084}                                   & 2022                                                                & Healthcare                                                             & \begin{tabular}[c]{@{}l@{}}Multi-hop permission delegation scheme using\\ PRE and ABE\end{tabular}                                                                               & Ethereum                                                               & N/A                                                                                                  \\ \hline
\citeauthor{10.3389/fmed.2022.837197} \cite{10.3389/fmed.2022.837197}                                     & 2022                                                                & Healthcare                                                             & \begin{tabular}[c]{@{}l@{}}Dynamic consent platform tailored for\\ clinical trials\end{tabular}                                                                                  & \begin{tabular}[c]{@{}c@{}}Hyperledger\\ Fabric\end{tabular}           & N/A                                                                                                  \\ \hline
\citeauthor{2022-crowmedII-consent} \cite{2022-crowmedII-consent}                                         & 2022                                                                & Healthcare                                                             & \begin{tabular}[c]{@{}l@{}}Health data management framework with\\ access rights\end{tabular}                                                                                    & Ethereum                                                               & N/A                                                                                                  \\ \hline
\citeauthor{2022-BDSS-blockchain-data-sharing} \cite{2022-BDSS-blockchain-data-sharing}                   & 2022                                                                & Healthcare                                                             & \begin{tabular}[c]{@{}l@{}}Framework with fine-grained access control and\\ permission revocation\end{tabular}                                                                   & N/A                                                                    & N/A                                                                                                  \\ \hline

\citeauthor{10.1145/3576842.3582379} \cite{10.1145/3576842.3582379} & 2023 & Internet of Things  &  \begin{tabular}[c]{@{}l@{}}Multi-phase consent mechanism for IoT data sharing, \\ using asymmetric cryptography\end{tabular} \\ \hline
\citeauthor{10.1016/j.jlamp.2023.100886} \cite{10.1016/j.jlamp.2023.100886}  & 2023 & Healthcare & \begin{tabular}[c]{@{}l@{}} System employing Event-B method, and  SmartDataTrust \\prototype\end{tabular}  \\ \hline
\citeauthor{10.1109/ICOIN56518.2023.10048920} \cite{10.1109/ICOIN56518.2023.10048920} & 2023 & Healthcare & \begin{tabular}[c]{@{}l@{}} A framework designed  to meet the requirements of the GDPR \end{tabular} \\ \hline
\end{tabular}
}
\end{table*}
\end{savenotes}


In general, smart contracts are widely used to manage data access permissions. The user can include rules inside the current contract state, and the data consumer can query only the allowed items. Regarding consent,  the user has the flexibility to allow or deny access by updating the contract state.

\subsection{Self-Sovereign Identity Management: A tool for building privacy-preserving applications}
\label{sec:ssidentity}

In the majority of systems, users depend on a central part to store their identities, including online services, healthcare, and government ID \cite{10.1109/ACCESS.2022.3216643}. On the other hand, self-sovereign identity gives the user more control over their identity information. Table \ref{tab:ssi-works-part-1} and \ref{tab:ssi-works-part-2} overviews the works discussed, including their publication year, the use case being examined, the key contributions made, the blockchain platform used, and whether the code is accessible. SSI has been explored for various use cases, including general-purpose identity management systems, passport-level attributes, biometric traits, the Internet of Things, open banking, public transportation, certificate management, healthcare, supply chain management, know-your-customer processes, and education institutions. Additionally, some works focus on enhancing privacy and security in SSI systems through different methods, such as fully homomorphic encryption, elliptic curve cryptography, and differential privacy.

\begin{itemize}
    \item \textbf{Biometric Traits and Identity Attributes:}
    \citeauthor{2019-ssi-digital-identity-bc} \cite{2019-ssi-digital-identity-bc} proposed a scheme to address identity fraud and thefts. The conceptual scheme combines biometric traits and identity attributes, allowing for retrospective verification of previous transactions.

    \item \textbf{Certificate Management and Cyber Threat Information:}
    \citeauthor{2021-an-innovative-did} \cite{2021-an-innovative-did} proposed a framework called Three Blockchains Identity Management with Elliptic Curve Cryptography (3BI-ECC) that allows users to self-generate and validate their own identities without depending on Trusted Third Parties (TTP).
    \citeauthor{2021-ssi-cyber-threat} presented a platform called Siddhi \cite{2021-ssi-cyber-threat}, a blockchain and SSI-enabled cyber threat information sharing platform that aims to improve traceability, reliability, privacy, scalability, anonymization, and data provenance of cyber threat information sharing among multiple organizations.
    
    \item \textbf{Higher Education Institutions:} \citeauthor{2022-hei-bct-ssi} proposed HEI-BCT \cite{2022-hei-bct-ssi}, a conceptual SSI framework to help students to manage their credentials in higher Education Institutions.
    
    \item \textbf{General-Purpose Identity Management Platforms:} Some authors have proposed general-purpose identity management platforms and solutions, such as uPort \cite{2020-uport-ims-ssi}, EverSSDI \cite{2019-everssdi}, BSelSovID \cite{2022-a-ssi-idm-bc}, SSIChain \cite{10.1109/ISCC53001.2021.9631518}, and distributed identity registries \cite{2021-dec-cross-net-idm}. These developments aim to provide decentralized and secure solutions for identity management, replacing centralized systems and trusted third parties.
    
    \item \textbf{Healthcare:} Several authors have proposed SSI solutions for healthcare, such as MediLinker \cite{2022-ssi-patient-centric-healthcare}, an SSI framework with privacy preservation \cite{2022-towards-ssi-privacy-fl}, Health-ID \cite{2021-health-id-did-healthcare}, and a solution with decentralized identity for data disclosure prevention \cite{2021-did-prevent-secondary}.
    
    \item \textbf{Industrial IoT:} Some other research has concentrated on implementing SSI in industrial settings, such as IIoT. \citeauthor{10.1109/LCN53696.2022.9843700} \cite{10.1109/LCN53696.2022.9843700} proposed a decentralized digital identity framework for IIoT devices, using the SSI model and implemented on Ethereum and Hyperledger Indy, that aims to avoid drawbacks of traditional approaches and enable autonomous communication.
    \citeauthor{2020-emission-vehicles-ssi} \cite{2020-emission-vehicles-ssi} proposed the integration of a public permissioned self-sovereign identities (SSI) framework with a permissioned consortium blockchain-based architecture for modern Internet of Things networks, specifically for vehicle networks, to improve security and access control concerning privacy for sensitive data.

\end{itemize}

Other studies have examined the use of SSI in different applications, which can be grouped into subcategories:

\begin{itemize}
    \item \textbf{Banking and KYC Processes:} \citeauthor{2022-ssi-kyc} \cite{2022-ssi-kyc} proposed an SSI-based framework focused on enhancing the know-your-customer (KYC) processes in the banking industry. Building upon this, \citeauthor{2020-bbm-ssi} \cite{2020-bbm-ssi} presented BBM, an SSI solution that aims to enhance the open banking experience by addressing customers' privacy concerns through the utilization of blockchain technology, demonstrating the versatility of SSI in addressing various banking challenges.

    \item \textbf{Identity Solutions for Passport-level Attributes and SSI Security:} \citeauthor{2018-bc-ssi} \cite{2018-bc-ssi} proposed a digital identity solution for passport-level attributes based on a generic provable claim collected through attestations of truth from third parties. \citeauthor{2020-enhancing-ssi} \cite{2020-enhancing-ssi} analyzed and proposed methods to improve the security and privacy of SSI systems in Hyperledger Indy, using an attribute sensitivity scoring model, mitigating man-in-the-middle attacks and determining the reputation of credential issuers.

    \item \textbf{Supply Chain Management in the Pharmaceutical and Transportation Sectors:} \citeauthor{10.15439/2022F194} \cite{10.15439/2022F194} presented an SSI approach for Inter-Organizational Business Processes (IOBP) and demonstrated its effectiveness by applying it to a pharmaceutical supply chain case study, implemented on the Ethereum Blockchain. In another domain, \citeauthor{2021-ssi-public-transport} \cite{2021-ssi-public-transport} investigated the potential for implementing SSI in the public transportation sector for cross-operator and cross-border travel, showcasing the versatility of SSI in enhancing passengers' overall efficiency and convenience across different sectors.

    \item \textbf{Comparing SSI Ecosystems and Food Supply Chain Traceability:} \citeauthor{10.1109/ACCESS.2022.3216643} \cite{10.1109/ACCESS.2022.3216643} examined identity management systems and their essential components, highlighting potential research gaps. Building on this, \citeauthor{10.1109/ISSE49799.2020.9272212} \cite{10.1109/ISSE49799.2020.9272212} proposed governing principles for evaluating any SSI ecosystem and compared the popular SSI ecosystems, uPort and Sovrin. Expanding the application of SSI to other domains, \citeauthor{2021-SSI-food-supply-chain} \cite{2021-SSI-food-supply-chain} suggested an identity management system utilizing SSI, blockchain, and IPFS technologies for transparency and traceability in the food supply chain.

\end{itemize}

\begin{savenotes}

\begin{table*}[ht!]
\tiny
\centering
\caption{(Part I) Summary of blockchain-based works with self-sovereign identity sorted by publication year, including information on the use case, key contribution, blockchain platform used, and software availability}
\label{tab:ssi-works-part-1}

\resizebox{\textwidth}{!}{

\begin{tabular}{lcclHZ}
\hline
\multicolumn{1}{c}{\textbf{Work}}                                                           & \textbf{\begin{tabular}[c]{@{}c@{}}Publication\\ Year\end{tabular}} & \textbf{\begin{tabular}[c]{@{}c@{}}Evaluation\\ Use Case\end{tabular}}         & \textbf{Key Contribution}                                                                                                                                       & \textbf{\begin{tabular}[c]{@{}c@{}}Blockchain platform\\ used by the authors\end{tabular}}        & \textbf{\begin{tabular}[c]{@{}c@{}}Software \\ Availability\end{tabular}}                         \\ \hline
\citeauthor{2018-bc-ssi} \cite{2018-bc-ssi}                                                 & 2018                                                                & \begin{tabular}[c]{@{}c@{}}Passport-level\\  attributes\end{tabular}           & SSI standard for passport-level attributes.                                                                                                                     & N/A                                                                            & N/A                                                                                               \\ \hline
\citeauthor{2019-everssdi} \cite{2019-everssdi}                                             & 2019                                                                & N/A                                                                            & \begin{tabular}[c]{@{}l@{}}SSI framework using smart contracts with \\ identity recovery using SNS and Ethereum Oracles.\end{tabular}                           & Ethereum                                                                       & N/A                                                                                               \\ \hline
\citeauthor{2019-ssi-digital-identity-bc} \cite{2019-ssi-digital-identity-bc}               & 2019                                                                & \begin{tabular}[c]{@{}c@{}}Biometric \\ Traits\end{tabular}                    & \begin{tabular}[c]{@{}l@{}}A dynamic SSI scheme that could vary or evolve\\ over time.\end{tabular}                                                             & N/A                                                                            & N/A                                                                                               \\ \hline
\citeauthor{10.1109/ACCESS.2019.2931173} \cite{10.1109/ACCESS.2019.2931173}                 & 2019                                                                & N/A                                                                            & \begin{tabular}[c]{@{}l@{}}A formal and mathematical model for understanding\\ the concept of SSI.\end{tabular}                                                 & N/A                                                                            & N/A                                                                                               \\ \hline
\citeauthor{2020-emission-vehicles-ssi} \cite{2020-emission-vehicles-ssi}                   & 2020                                                                & Internet of Things                                                             & SSI architecture for IoT and Smart Vehicles.                                                                                                                    & Hyperledger Indy                                                               & N/A                                                                                               \\ \hline
\citeauthor{2020-uport-ims-ssi} \cite{2020-uport-ims-ssi}                                   & 2020                                                                & N/A                                                                            & Open-source identity management using smart contracts.                                                                                                          & Ethereum                                                                       & N/A                                                                                               \\ \hline
\citeauthor{10.1109/MobileCloud48802.2020.00021} \cite{10.1109/MobileCloud48802.2020.00021} & 2020                                                                & N/A                                                                            & Specifications for evaluating SSI solutions.                                                                                                                    & N/A                                                                            & N/A                                                                                               \\ \hline
\citeauthor{10.1109/ISSE49799.2020.9272212} \cite{10.1109/ISSE49799.2020.9272212}           & 2020                                                                & \begin{tabular}[c]{@{}c@{}}uPort and Sovrin \\ ecosystems\end{tabular}         & \begin{tabular}[c]{@{}l@{}}Governing principles of SSI for analyzing the\\ effectiveness of SSI ecosystems.\end{tabular}                                        & N/A                                                                            & N/A                                                                                               \\ \hline
\citeauthor{2020-bbm-ssi} \cite{2020-bbm-ssi}                                               & 2020                                                                & Open Banking                                                                   & SSI management for open banking.                                                                                                                                & N/A                                                                            & N/A                                                                                               \\ \hline
\citeauthor{2020-enhancing-ssi} \cite{2020-enhancing-ssi}                                   & 2020                                                                & \begin{tabular}[c]{@{}c@{}}Personal Data\\ Disclosure\end{tabular}             & \begin{tabular}[c]{@{}l@{}}SSI model for attribute sensitivity in credential \\ exchanges.\end{tabular}                                                         & Hyperledger Indy                                                               & N/A                                                                                               \\ \hline
\citeauthor{2021-ssi-public-transport} \cite{2021-ssi-public-transport}                     & 2021                                                                & \begin{tabular}[c]{@{}c@{}}Public \\ Transportation\end{tabular}               & SSI model for multiple credentials for travel.                                                                                                                  & Hyperledger Indy                                                               & N/A                                                                                               \\ \hline
\citeauthor{2021-an-innovative-did} \cite{2021-an-innovative-did}                           & 2021                                                                & \begin{tabular}[c]{@{}c@{}}Certificate \\ Management\end{tabular}              & \begin{tabular}[c]{@{}l@{}}SSI management framework with Elliptic Curve\\ Cryptography.\end{tabular}                                                            & N/A                                                                            & N/A                                                                                               \\ \hline
\citeauthor{2021-did-prevent-secondary} \cite{2021-did-prevent-secondary}                   & 2021                                                                & Healthcare                                                                     & \begin{tabular}[c]{@{}l@{}}Decentralized Identity System with \\ Cheon-Kim-Kim-Song (CKKS) fully homomorphic\\ encryption to protect user privacy.\end{tabular} & Hyperledger Indy                                                               & N/A                                                                                               \\ \hline
\citeauthor{2021-health-id-did-healthcare}\footnote{\url{https://github.com/ibrahimtariqjaved/healthid}} \cite{2021-health-id-did-healthcare}             & 2021                                                                & Healthcare                                                                     & \begin{tabular}[c]{@{}l@{}}Patients Identity management model using\\ smart contracts.\end{tabular}                                                             & Ethereum                                                                       & \textit{\begin{tabular}[c]{@{}c@{}}https://github.com/ibrahimtariqjaved/\\ healthid\end{tabular}} \\ \hline
\citeauthor{2021-dec-cross-net-idm} \cite{2021-dec-cross-net-idm}                           & 2021                                                                & N/A                                                                            & \begin{tabular}[c]{@{}l@{}}A DLT-agnostic architecture and protocols \\ for identity management.\end{tabular}                                                   & Hyperledger Indy                                                               & N/A                                                                                               \\ \hline
\citeauthor{2021-SSI-food-supply-chain} \cite{2021-SSI-food-supply-chain}             & 2021                                                                & Supply Chain                                                                   & \begin{tabular}[c]{@{}l@{}}An SSI management system for process and certifications \\ in the food supply chain.\end{tabular}                                    & Ethereum                                                                       & N/A                                                                                               \\ \hline
                                                                                
\end{tabular}

}

\end{table*}

\end{savenotes}

\begin{savenotes}

\begin{table*}[ht!]
\tiny
\centering
\caption{(Part II) - Summary of blockchain-based works with self-sovereign identity sorted by publication year, including information on the use case, key contribution, blockchain platform used, and software availability}
\label{tab:ssi-works-part-2}

\resizebox{\textwidth}{!}{

\begin{tabular}{lcclHZ}
\hline
\multicolumn{1}{c}{\textbf{Work}}                                                           & \textbf{\begin{tabular}[c]{@{}c@{}}Publication\\ Year\end{tabular}} & \textbf{\begin{tabular}[c]{@{}c@{}}Evaluation\\ Use Case\end{tabular}}         & \textbf{Key Contribution}                                                                                                                                       & \textbf{\begin{tabular}[c]{@{}c@{}}Blockchain platform\\ used by the authors\end{tabular}}        & \textbf{\begin{tabular}[c]{@{}c@{}}Software \\ Availability\end{tabular}}                         \\ \hline

\citeauthor{10.1109/ISCC53001.2021.9631518} \cite{10.1109/ISCC53001.2021.9631518}           & 2021                                                                & N/A                                                                            & \begin{tabular}[c]{@{}l@{}}An SSI platform to allow users to control\\ their identities autonomously.\end{tabular}                                              & Hyperledger Fabric                                                             & N/A                                                                                               \\ \hline
\citeauthor{2021-ssi-cyber-threat} \cite{2021-ssi-cyber-threat}                             & 2021                                                                & \begin{tabular}[c]{@{}c@{}}Cyber Threat \\ Information\end{tabular}            & An SSI platform to share Cyber threat information.                                                                                                              & \begin{tabular}[c]{@{}c@{}}Rahasak with “Aplos” \\ smart contract\end{tabular} & N/A                                                                                               \\ \hline
\citeauthor{2022-ssi-kyc} \cite{2022-ssi-kyc}                                               & 2022                                                                & KYC                                                                            & An SSI-based framework for KYC process.                                                                                                                         & N/A                                                                            & N/A                                                                                               \\ \hline
\citeauthor{2022-a-ssi-idm-bc} \cite{2022-a-ssi-idm-bc}                                     & 2022                                                                & N/A                                                                            & \begin{tabular}[c]{@{}l@{}}An SSI management system with data encryption\\ combining blockchain and IPFS.\end{tabular}                                          & N/A                                                                            & N/A                                                                                               \\ \hline
\citeauthor{2022-ssi-patient-centric-healthcare} \cite{2022-ssi-patient-centric-healthcare} & 2022                                                                & Healthcare                                                                     & An SSI patient-centric healthcare identity wallet.                                                                                                              & Hyperledger Indy                                                               & N/A                                                                                               \\ \hline
\citeauthor{2022-towards-ssi-privacy-fl} \cite{2022-towards-ssi-privacy-fl}                 & 2022                                                                & Healthcare                                                                     & \begin{tabular}[c]{@{}l@{}}An SSI framework with privacy protection using\\ differential privacy and federated learning.\end{tabular}                           & N/A                                                                            & N/A                                                                                               \\ \hline
\citeauthor{10.1109/LCN53696.2022.9843700} \cite{10.1109/LCN53696.2022.9843700}             & 2022                                                                & \begin{tabular}[c]{@{}c@{}}Industrial\\ Internet-of-Things (IIoT)\end{tabular} & A decentralized identity management framework for IIoT.                                                                                                         & \begin{tabular}[c]{@{}c@{}}Ethereum and \\ Hyperledger Indy\end{tabular}       & N/A                                                                                               \\ \hline
\citeauthor{2022-hei-bct-ssi} \cite{2022-hei-bct-ssi}                                       & 2022                                                                & \begin{tabular}[c]{@{}c@{}}Education \\ Institutions\end{tabular}              & An SSI management framework for student credentials.                                                                                                            & N/A                                                                            & N/A                                                                                               \\ \hline
\citeauthor{10.15439/2022F194} \cite{10.15439/2022F194}                                     & 2022                                                                & \begin{tabular}[c]{@{}c@{}}Pharmaceutical \\ Supply Chain\end{tabular}         & \begin{tabular}[c]{@{}l@{}}An SSI approach for Inter-Organizational Business \\ Processes (IOBP).\end{tabular}                                                  & Ethereum                                                                       & N/A     \\ \hline

\end{tabular}

}

\end{table*}

\end{savenotes}


\section{Blockchain platforms with a focus on privacy}
\label{sec:platforms-with-focus-on-privacy}

This section introduces some platforms that provide privacy for transaction details. 
Table \ref{tab:privacy-platforms} compares the privacy platforms in terms of native cryptocurrency and token, privacy technique used, limitations, and open-source software availability. Additionally, Table \ref{tab:privacy-techniques-layers} presents a systematic categorization of privacy techniques employed across three separate layers in blockchain technology: network (Layer-0), on-chain (Layer-1), and off-chain (Layer-2). Layer-0 emphasizes privacy at the network level, Layer-1 techniques are integrated within the blockchain protocol, and Layer-2 represents an off-chain strategy.

\begin{savenotes}
\begin{table*}[ht!]
\centering
\caption{blockchain-based protocols with a focus on privacy}
\setlength{\tabcolsep}{0.8em}
\def\arraystretch{1.1}
\label{tab:privacy-platforms}

\resizebox{0.9\textwidth}{!}{

\begin{tabular}{lccclZ}
\hline
\multicolumn{1}{c}{\textbf{Protocol}} &
  \textbf{\begin{tabular}[c]{@{}c@{}}Launch \\ Year\end{tabular}} &
  \textbf{\begin{tabular}[c]{@{}c@{}}Native\\ Cryptocurrency/Token\end{tabular}} &
  \textbf{\begin{tabular}[c]{@{}c@{}}Privacy \\ Technique\end{tabular}} &
  \multicolumn{1}{c}{\textbf{Limitations}} &
  \textbf{\begin{tabular}[c]{@{}c@{}}Open-Source\\ Software\end{tabular}} \\ \hline
Zcash\footnote{\url{https://github.com/zcash}} &
  2016 &
  Yes (ZEC) &
  zk-SNARKs &
  \begin{tabular}[c]{@{}l@{}}- Initial trusted setup;\\ - No smart contracts support;\\ - High energy consumption;\\ - Slow transaction processing;\end{tabular} &
  \textit{\begin{tabular}[c]{@{}c@{}}https://github.com/\\ zcash\end{tabular}} \\ \hline
\begin{tabular}[c]{@{}l@{}}Tornado\\ Cash\footnote{\url{https://development.tornadocash.community/tornadocash}}\end{tabular} &
  2019 &
  No &
  \begin{tabular}[c]{@{}c@{}}zk-SNARKs and\\  Ethereum mixer\end{tabular} &
  \begin{tabular}[c]{@{}l@{}}- Initial trusted setup;\\ - Limited to Ethereum;\\ - Users may face high \\ transaction costs;\end{tabular} &
  \textit{\begin{tabular}[c]{@{}c@{}}https://development.\\ tornadocash.community/\\ tornadocash\end{tabular}} \\ \hline
Incognito\footnote{\url{https://github.com/incognitochain}} &
  2019 &
  Yes (PRV) &
  \begin{tabular}[c]{@{}c@{}}ZKP, Ring Signatures, \\ and Homomorphic \\ Commitment\end{tabular} &
  - No smart contracts support; &
  \textit{\begin{tabular}[c]{@{}c@{}}https://github.com/\\ incognitochain\end{tabular}} \\ \hline
\begin{tabular}[c]{@{}l@{}}Aztec\\ Network\footnote{\url{https://github.com/AztecProtocol}}\end{tabular} &
  2020 &
  No &
  zk-SNARKs &
  \begin{tabular}[c]{@{}l@{}}- Initial trusted setup;\\ - Limited to Ethereum;\\ - zk-SNARKs and rollups\\ techniques, which can be\\ complex operations;\end{tabular} &
  \textit{\begin{tabular}[c]{@{}c@{}}https://github.com/\\ AztecProtocol/\end{tabular}} \\ \hline
Monero\footnote{\url{https://github.com/monero-project}} &
  2014 &
  Yes (XMR) &
  \begin{tabular}[c]{@{}c@{}}Ring Signatures,\\  Confidential \\ Transactions, and \\ Stealth Addresses\end{tabular} &
  \begin{tabular}[c]{@{}l@{}}- No smart contracts support;\\ - High energy consumption;\\ - Slow transaction processing;\end{tabular} &
  \textit{\begin{tabular}[c]{@{}c@{}}https://github.com/\\ monero-project\end{tabular}} \\ \hline
\begin{tabular}[c]{@{}l@{}}Secret\\ Network\footnote{\url{https://github.com/scrtlabs}}\end{tabular} &
  2020 &
  \begin{tabular}[c]{@{}c@{}}Yes (SCRT)\\ (Token)\end{tabular} &
  \begin{tabular}[c]{@{}c@{}}Trusted Execution \\ Environments and SMPC\end{tabular} &
  \begin{tabular}[c]{@{}l@{}}- Needs a trusted execution \\ environment hardware;\end{tabular} &
  \textit{\begin{tabular}[c]{@{}c@{}}https://github.com/\\ scrtlabs\end{tabular}} \\ \hline
Grin\footnote{\url{https://github.com/mimblewimble/grin}} &
  2019 &
  Yes (GRIN) &
  \begin{tabular}[c]{@{}c@{}}Mimblewimble\\  protocol\end{tabular} &
  \begin{tabular}[c]{@{}l@{}}- Security concerns;\\ - Lack of adoption;\\ - High energy consumption; \\ - Slow transaction processing;\end{tabular} &
  \textit{\begin{tabular}[c]{@{}c@{}}https://github.com/\\ mimblewimble/grin\end{tabular}} \\ \hline
Beam\footnote{\url{https://github.com/BeamMW/beam}} &
  2019 &
  Yes (BEAM) &
  \begin{tabular}[c]{@{}c@{}}Mimblewimble\\  protocol\end{tabular} &
  \begin{tabular}[c]{@{}l@{}}- Security concerns;\\ - Lack of adoption;\\ - High energy consumption; \\ - Slow transaction processing;\end{tabular} &
  \textit{\begin{tabular}[c]{@{}c@{}}https://github.com/\\ BeamMW/beam\end{tabular}} \\ \hline
\end{tabular}

}

\end{table*}
\end{savenotes}

\begin{table*}[ht!]

\centering

\caption{Comparative analysis of zero-knowledge proof systems: zk-SNARKs, zk-STARKs, and Bulletproofs \cite{awesome_zkp}.}

\label{tab:zkp-systems}

\resizebox{\textwidth}{!}{

\begin{tabular}{lcccccc}
\hline
\textbf{ZKP System}                        & \textbf{\begin{tabular}[c]{@{}c@{}}Algorithmic\\ complexity: prover\end{tabular}} & \textbf{\begin{tabular}[c]{@{}c@{}}Algorithmic \\ complexity: verifier\end{tabular}} & \textbf{\begin{tabular}[c]{@{}c@{}}Communication\\ complexity (proof size)\end{tabular}} & \textbf{\begin{tabular}[c]{@{}c@{}}Trusted setup \\ required?\end{tabular}} & \textbf{\begin{tabular}[c]{@{}c@{}}Post-quantum \\ secure\end{tabular}} & \textbf{\begin{tabular}[c]{@{}c@{}}Blockchain \\ Platforms\end{tabular}}        \\ \hline
zk-SNARKs \cite{10.5555/2671225.2671275}   & $ O(N * log(N)) $                                                                 & $ ~O(1) $                                                                            & $ ~O(1) $                                                                                & Yes                                                                         & No                                                                      & \begin{tabular}[c]{@{}c@{}}Zcash, \\ Tornado Cash,\\ Aztec Network\end{tabular} \\ \hline
zk-STARKs \cite{cryptoeprint:2018/046}     & $ O(N * poly-log(N))$                                                             & $ O(poly-log(N))$                                                                    & $ O(poly-log(N))$                                                                        & No                                                                          & Yes                                                                     & -                                                                               \\ \hline
Bulletproofs \cite{sikoba2019bulletproofs} & $ O(N * log(N)) $                                                                 & $ O(N) $                                                                             & $ O(log(N)) $                                                                            & No                                                                          & No                                                                      & Monero                                                                          \\ \hline
\end{tabular}

}

\end{table*}

\begin{table}[ht!]
\tiny
\centering
\caption{Comparative Analysis of Privacy Techniques Employed at Different Blockchain Layers: Layer-0 (Network), Layer-1 (On-chain), and Layer-2 (Off-chain)}
\label{tab:privacy-techniques-layers}

\resizebox{0.6\textwidth}{!}{

\begin{tabular}{ll}
\hline
\textbf{Layer}                                                & \textbf{Privacy Techniques}                                                                            \\ \hline
\begin{tabular}[c]{@{}l@{}}Layer-2\\ (Off-chain)\end{tabular} & \begin{tabular}[c]{@{}l@{}}zk-SNARKs, \\ zk-STARKs,\\ and Bulletproofs.\end{tabular}                   \\ \hline
\begin{tabular}[c]{@{}l@{}}Layer-1\\ (On-chain)\end{tabular} &
  \begin{tabular}[c]{@{}l@{}}Indistinguishability Obfuscation (IO) \cite{Garg2016}, \\ Homomorphic encryption \cite{homenc}, \\ Secure multiparty computation \cite{zyskind2015enigma},\\ Trusted hardware-assisted \cite{8806762},\\ Mixing protocols \cite{8823022},\\ CoinSwap (Atomic Swap) \cite{Deshpande2020},\\ Dandelion \cite{2017dandelion, fanti2018dandelion},\\ Confidential transactions (Pedersen commitments),\\ TumbleBit \cite{Liu2019},\\ MimbleWimble \cite{mimblewimble-white-paper},\\ Anonymous signatures, and\\ Attribute-based encryption.\end{tabular} \\ \hline
\begin{tabular}[c]{@{}l@{}}Layer-0\\ (Network)\end{tabular}   & \begin{tabular}[c]{@{}l@{}}Invisible Internet Project (I2P) and \\ The Onion Router (TOR)\end{tabular} \\ \hline
\end{tabular}
}

\end{table}

\section{Identity management platforms}
\label{sec:identity-management-platforms}

This section presents an overview of some decentralized platforms that can be used to build identity solutions, including Serto, Veramo, Sovrin Network, Civic, Evernym, and Identity.
Table \ref{tab:ssi-platforms} summarizes the self-sovereign identity platforms in terms of the blockchain platform used, smart contract capabilities, limitations, and open-source software availability.



\begin{savenotes}
\begin{table*}[ht!]
\centering
\caption{Blockchain-based platforms with a focus on identity management}
\label{tab:ssi-platforms}

\resizebox{0.9\textwidth}{!}{

\begin{tabular}{lccclZ}
\hline
\textbf{\begin{tabular}[c]{@{}l@{}}Blockchain-based \\ Platform\end{tabular}} & \textbf{\begin{tabular}[c]{@{}c@{}}Launch\\ Year\end{tabular}} & \textbf{\begin{tabular}[c]{@{}c@{}}Blockchain\\ Platform\end{tabular}}    & \textbf{\begin{tabular}[c]{@{}c@{}}Smart Contract \\ Capabilities\end{tabular}} & \textbf{Limitations}                                                                                                              & \textbf{\begin{tabular}[c]{@{}c@{}}Open-Source\\ Software\end{tabular}}                  \\ \hline
Serto\footnote{\url{https://github.com/sertoid}}                                                                         & 2021                                                           & N/A                                                                       & No                                                                              & \begin{tabular}[c]{@{}l@{}}- No native blockchain integration;\\ - The agent must be deployed in \\ a cloud service;\end{tabular} & \textit{\begin{tabular}[c]{@{}c@{}}https://github.com/\\ sertoid\end{tabular}}           \\ \hline
Veramo\footnote{\url{https://github.com/veramolabs}}                                                                        & 2021                                                           & N/A                                                                       & No                                                                              & \begin{tabular}[c]{@{}l@{}}- No native blockchain integration;\\ - The agent must be deployed in \\ a cloud service;\end{tabular} & \textit{\begin{tabular}[c]{@{}c@{}}https://github.com/\\ veramolabs\end{tabular}}        \\ \hline
Sovrin Network\footnote{\url{https://github.com/sovrin-foundation}}                                                                & 2016                                                           & Hyperledger Indy                                                          & Yes                                                                             & \begin{tabular}[c]{@{}l@{}}- Hyperledger Indy specific;\\ - Less interoperability with other \\ blockchains;\end{tabular}         & \textit{\begin{tabular}[c]{@{}c@{}}https://github.com/\\ sovrin-foundation\end{tabular}} \\ \hline
Civic\footnote{\url{https://github.com/civicteam}}                                                                         & 2017                                                           & \begin{tabular}[c]{@{}c@{}}Ethereum and\\ Solana Integration\end{tabular} & Yes                                                                             & \begin{tabular}[c]{@{}l@{}}- Focused on Ethereum and Solana\\ networks;\end{tabular}                                              & \textit{\begin{tabular}[c]{@{}c@{}}https://github.com/\\ civicteam\end{tabular}}         \\ \hline
Evernym\footnote{\url{https://gitlab.com/evernym}}                                                                       & 2013                                                           & Hyperledger Indy                                                          & Yes                                                                             & - Limited to Hyperledger Indy;                                                                                                    & \textit{\begin{tabular}[c]{@{}c@{}}https://gitlab.com/\\ evernym\end{tabular}}           \\ \hline
Identity.com\footnote{\url{https://github.com/identity-com}}                                                                  & 2020                                                           & Solana                                                                    & Yes                                                                             & - Limited to Solana specification;                                                                                                & \textit{\begin{tabular}[c]{@{}c@{}}https://github.com/\\ identity-com\end{tabular}}      \\ \hline
\end{tabular}

}

\end{table*}
\end{savenotes}

\section{Research Opportunities}
\label{sec:research-opportunities}

There are many opportunities for research in user privacy, self-sovereign identity, and machine learning facilitated through blockchain technology. In particular, we highlight that these areas require further improvement with respect to the privacy and security of users' personal information.

\subsection{Privacy and Consent Management}

Some research opportunities for privacy-enhancing mechanisms with blockchain technology include the design and evaluation of the following approaches:

\begin{itemize}
    \item Encryption including proxy re-encryption (PRE), homomorphic encryption, and attribute-based encryption (ABE). Additionally, evaluate the performance of a system to perform encryption operations on-chain using smart contracts instead of off-chain.
    \item NTRU post-quantum cryptographic scheme for IoT edge devices and sensitive applications.
    \item The feasibility of differential privacy when compared with encryption approaches;
\end{itemize}

Different use cases can be explored to evaluate these protocols, including financial services and the Internet of Medical Things (IoMT). These applications generally request personal information and create sensitive records. Moreover, the system must provide user consent to share specific attributes with other organizations. For applications with the sharing of large file sizes,  a hybrid scheme combining on-chain and a peer-to-peer off-chain, such as InterPlanetary File System (IPFS), can allow the sharing of different file types while protecting user privacy \cite{JAYABALAN2022152, Kumar2021}. Other privacy requirements involve the user revoking or removing access to personal data. It allows the development of privacy-preserving systems in compliance with privacy regulations.

\subsubsection{Opportunities in development of privacy-focused platforms}

As presented in Section \ref{sec:platforms-with-focus-on-privacy}, some blockchain platforms focus on preserving the privacy of the sender, the receiver, and the amount in a transaction. However, some opportunities in the development of blockchain-based platforms should be considered:

\begin{itemize}
    \item \textbf{Smart Contract Capabilities:} Nowadays, different applications can employ smart contracts to automate tasks while protecting privacy in different industrial use cases, including the financial sector, the Internet of Things, and healthcare.

    \item \textbf{Cross-chain transactions:} Allowing privacy-focused platforms to interact with other blockchains and systems.
     
    \item \textbf{Confidential Assets:} Creating digital assets that can be transferred securely without revealing their type, quantity, or other sensitive information is an opportunity to increase privacy for users. 

    \item \textbf{Decentralized Identity Management:} Decentralized identity solutions can empower users to have control over their personal data and credentials. This could lead to applications such as identity verification, access control, and data sharing.
    
    \item \textbf{Regulatory Compliance:} Addressing regulatory concerns and ensuring compliance with anti-money laundering (AML) and know-your-customer (KYC) requirements is crucial for privacy platforms.

    \item \textbf{Confidential IP:} Addresses the disclosure of IP addresses during transaction initiation. Confidential IP as presented in Table\ref{tab:privacy-techniques-layers}, is a network protocol (Layer-0) to provide privacy.
\end{itemize}

\subsection{Self-Sovereign Identity and Verifiable Credentials}

The security and interoperability of identity management systems are challenging research topics. Some potential opportunities include:

\begin{itemize}
    \item Evaluate different settings to implement SSI protocols, including healthcare \cite{10.1109/ACCESS.2020.2994090}, market \cite{Feulner2022}, and academic \cite{9770913};

    \item Investigate how the SSI systems could be integrated with Short Group Signatures proposed by \citeauthor{10.1007/978-3-540-28628-8_3} \cite{10.1007/978-3-540-28628-8_3} formerly BBS signatures. Existing implementations proposed by the Decentralized Identity Foundation include zero-knowledge proof integration \cite{repo-mattr}. It enables selective sharing in a verifiable credentials format;

    \item Evaluate the interoperability and scalability of identity management systems when compared with the existing centralized approaches;

\end{itemize}

\subsection{Scalability}

Despite the blockchain features, one research topic is to evaluate the scalability regarding throughput and latency performance of privacy-preserving solutions integrated with identity management and machine learning algorithms. Additionally, different consensus algorithms can be proposed and evaluated concerning operational costs in use cases with sensitive data. For instance, \citeauthor{2022-survey-consensus-algorithm} \cite{2022-survey-consensus-algorithm} and \citeauthor{BAO2023111555} \cite{BAO2023111555} analyzed different aspects of consensus mechanisms such as security and challenges. Similarly, \citeauthor{SANKA2021103232} \cite{SANKA2021103232} reviewed the scalability issues in blockchain, including transactions per second (TPS) and data storage. \cite{10.1109/ACCESS.2021.3117662} examined the scalability of IoMT systems to provide design recommendations in blockchain-based medical things. In the context of the industrial sector, there are some scalability challenges that can be evaluated:

\begin{itemize}
    \item \textbf{The IoT devices addition:} evaluate system performance with the addition of new devices to create transactions, such as sensor nodes and medical devices.

    \item \textbf{On-chain and Off-chain Storage:} examine how the system handles the storage of an increased number of transactions created by new users and devices. Alternatively, evaluate how off-chain storage such as IPFS improves the system scalability.

\end{itemize}

In general, scalability is a challenge in blockchain systems. Some solutions aim to address this issue by enhancing the blockchain protocol at the layer-1 level, such as changing block size and consensus mechanisms. Meanwhile, alternative protocols focus on off-chain approaches, operating within the layer-2 domain \cite{8962150}.

\section{Conclusion}
\label{sec:conclusion}

The increased data sharing in modern systems raises concerns about user privacy and how various entities collect, store, use, and protect data. This survey examined strategies and solutions for privacy protection, consent, and identity management in multi-stakeholder settings. Firstly we analyzed different schemes that combine blockchain and privacy in the supply chain, healthcare, and the Internet of Things application domains. 
Some commonly used privacy techniques include encryption-based systems and zero-knowledge proofs, but these schemes tend to rely on external entities and trusted third parties to protect and process sensitive information, limiting privacy.  Additionally, we analyzed different existing blockchain platforms with built-in support for privacy protocols. Correlated with privacy, we analyzed user consent and identity management studies as long as existing platforms enable users to control their identity using a decentralized infrastructure. Our survey indicates that there are promising approaches for privacy in blockchain-based applications, but more research is needed to make them practical for real-world applications. Besides, the reliance on trusted third parties must be minimized to enhance trust and privacy.

\section{Acknowledgments}

This work was partly funded by the São Paulo Research Foundation (FAPESP), grants 2021/10921-2 and 2013/07375-0.


\bibliographystyle{IEEEtranN}
\bibliography{related_surveys,references,consent_works,privacy_works,did_ssi_works,machine_learning_works,new-works-analysis}


\end{document}